\documentclass[]{aa}

\usepackage[varg]{txfonts}
\usepackage{graphicx}

\graphicspath{{./}{figs/}}

\begin{document}

\title{Cepheids with giant companions. II.} 
\subtitle{Spectroscopic confirmation of nine new double-lined binary systems composed of two Cepheids
  \thanks{Based on observations collected at the European Southern Observatory, Chile}\fnmsep
  \thanks{This paper includes data gathered with the 6.5m Magellan Clay Telescope at Las Campanas Observatory, Chile.}
  } 

\author{Bogumi{\l} Pilecki\inst{1}
  \and Ian B. Thompson\inst{2}
  \and Felipe Espinoza-Arancibia\inst{1}
  \and Gergely Hajdu\inst{1}
  \and Wolfgang Gieren\inst{3}
  \and M{\'o}nica Taormina\inst{1}
  \and Grzegorz Pietrzy{\'n}ski\inst{1}
  \and Weronika Narloch\inst{1}
  \and Giuseppe Bono\inst{4}
  \and Alexandre Gallenne\inst{5,6}
  \and Pierre Kervella\inst{7}
  \and Piotr Wielg{\'o}rski\inst{1}
  \and Bart{\l}omiej Zgirski\inst{3}
  \and Dariusz Graczyk\inst{8}
  \and Paulina Karczmarek\inst{3}
  \and Nancy R. Evans\inst{9}
  } 

\institute{Centrum Astronomiczne im. Miko{\l}aja Kopernika, PAN, Bartycka 18, 00-716 Warsaw, Poland\\
  \email{pilecki@camk.edu.pl}
  \and Carnegie Observatories, 813 Santa Barbara Street, Pasadena, CA 91101-1292, USA
  \and Universidad de Concepci{\'o}n, Departamento de Astronom{\'i}a, Casilla 160-C, Concepci{\'o}n, Chile
  \and Dipartimento di Fisica Universit`a di Roma Tor Vergata, viadella Ricerca Scientifica 1, 00133 Rome, Italy,
  \and Instituto de Astrofísica, Universidad Andrés Bello, Fernández Concha 700, Las Condes, Santiago, Chile
  \and French-Chilean Laboratory for Astronomy, IRL 3386, CNRS, Casilla 36-D, Santiago, Chile
  \and LESIA, Observatoire de Paris, PSL, CNRS, UPMC, Univ. Paris-Diderot, 5 place Jules Janssen, 92195 Meudon, France
  \and Centrum Astronomiczne im. Miko{\l}aja Kopernika, PAN, Rabia{\'n}ska 8, 87-100 Toru{\'n}, Poland,
  \and Smithsonian Astrophysical Observatory, MS 4, 60 Garden St., Cambridge, MA 02138
  }

\date{Submitted 31 December 2023 / Accepted 14 March 2024}

\abstract
{
Binary Cepheids with giant companions are crucial for studying the physical properties of Cepheid variables, in particular providing the best means to measure their masses. Systems composed of two Cepheids are even more important but to date, only one such system in the Large Magellanic Cloud (LMC) was known.}
{
Our current aim is to increase the number of these systems tenfold and provide their basic characteristics. The final goal is to obtain the physical properties of the component Cepheids, including their masses and radii, and to learn about their evolution in the multiple systems, also revealing their origin.}
{
We started a spectroscopic monitoring of nine unresolved pairs of Cepheids from the OGLE catalog, to check if they are gravitationally bound. Two of these so-called double Cepheids are located in the LMC, five in the Small Magellanic Cloud (SMC), and two in the Milky Way (MW).}
{
We report the spectroscopic detection of binarity of all nine of these double Cepheids with orbital periods ranging from 2 to 18 years. This increases the number of known {\it binary double} (BIND) Cepheids from 1 to 10 and triples the number of all confirmed {\it double-lined binary} (SB2) Cepheids. For five BIND Cepheids disentangled pulsational light curves of the components show anti-correlated phase shifts due to orbital motion. We show the first empirical evidence that typical period-luminosity relations (PLRs) are rather {\it binary Cepheid} PLRs, as they include the companion's light.
}
{
The statistics of pulsation period ratios of BIND Cepheids do not agree with those expected for pairs of the same-age Cepheids. These ratios together with the determined mass ratios far from unity suggest merger-origin of at least one component for about half of the systems. The SMC and MW objects are the first found in SB2 systems composed of giants in their host galaxies. The Milky Way BIND Cepheids are also the closest such systems, being located at about 11 and 26 kpc.
}


\maketitle

\section{Introduction} \label{sec:intro}

Classical Cepheids (hereafter also Cepheids) form probably the most important class of pulsating stars. Because of the period-luminosity relation they obey, they are important distance indicators in the local Universe, providing a fundamental step of the cosmic distance ladder and connecting our Milky Way galaxy to galaxies in the Local Group and beyond. They are also key objects for testing the predictions of stellar evolution and stellar pulsation theories. There are almost 15000 classical Cepheids known \citep{Pietruk_2021_CEP_num_MW_total}, mostly in the Milky Way (\citealt{Pojmanski_2005_ASAS_cat5} and references therein, \citealt{Soszynski_2020AcA_upd_MW_ceps}) and Magellanic Clouds \citep{Soszynski_2017AcA_OCVS_MC_Cep}. Even 80\% of these Cepheids are expected to be members of binary systems \citep{Kervella_2019_Multiplicity}. However, only about 0.1\% of them were found in eclipsing binary systems, and even less in double-lined spectroscopic binaries, while both these features {may be used to provide indispensable information about the physical properties of Cepheids.

Our study of Cepheids in eclipsing binary systems have already brought a wealth of data regarding the physical properties of Cepheids, with the accuracy of 0.5-2\% for the most interesting Cepheid masses and radii \citep[e.g.,][]{cep227nature2010, cep2532apj2015, allcep_pilecki_2018}. During that study several interesting cases were identified, including an extremely rare system composed of two Cepheids, OGLE-LMC-CEP-1718 \citep{cep1718apj2014}. Having two Cepheids in one system put important constraints on models as their ages have to be equal, and physical parameters similar and related to their pulsation periods. We can also assume they were born with the same chemical composition. Moreover, a subsequent study of OGLE-LMC-CEP-1718 by \citet{allcep_pilecki_2018} showed that the more evolutionary advanced component is actually slightly less massive, which may be an important hint in the ongoing discussion on the origin of Cepheid mass discrepancy problem \citep{Cassisi_Salaris_2011_ApJ, Anderson_2016_Rotation}.

Although first unresolved pairs of Cepheids (dubbed double Cepheids) were identified almost 30 years ago \citep{Alcock_1995_double_cepheids}\footnote{We ignore here a pair of Cepheids, CE Cas AB, which was called double Cepheid \citep{Sandage_Tammann_1969_doubleCep_CE_Cas} but forms a visual pair separated by 2.5".}, to date only one was spectroscopically confirmed as a binary system and analyzed, which makes any statistical analysis impossible. A very limited parameter space may also render any theoretical study inconclusive. A larger set of such stars occupying wider parameter space would provide an enormous insight into the pulsation and evolution theory through a comparative analysis of the differences between the components. All models would have to predict correct mass and period ratios (together with other observables) of the same-age and similar-composition components for several systems at the same time.

The problem with the spectroscopic confirmation of binarity of double Cepheids was their relative faintness and long expected orbital periods. One of the aims of our project is to confirm and monitor a statistically significant sample of such systems in different environments and provide their basic orbital and physical parameters. Such a sample will form a base for many future studies, including comparisons with model predictions of evolution and pulsation theories.

It is important to note, that such systems will always be double-lined spectroscopic binaries (SB2) in visual bands, as their Cepheid components are both giant stars with similar brightness and atmospheric properties.  A reliable mass determination (using Kepler's laws) is possible practically only for SB2 systems, but until recently only five such double-lined binary systems containing Cepheids were known \citep{allcep_pilecki_2018}. The rest of systems with Cepheids are single-lined binaries (SB1), which makes mass estimates for their components very uncertain, with typical accuracy of 10-20\% \citep{Evans_2018_V350SGr_mass}. For a few of them a great effort is taken to observe them from space in far UV, where companion's lines are detectable and velocities can be measured \citep{Gallenne_2018_V1334Cyg_Cep}. In this paper we will keep calling {\it a spectroscopic double-lined binary} (or SB2) only objects for which lines of both components are identified in the same spectrum, at the same wavelength range.

In the first paper of the series \citep[hereafter P21]{cepgiant1_2021} we proposed a new method of identification of Cepheids in SB2 systems, in which Cepheids that are too bright for their periods, have similar or redder color and lower pulsation amplitudes are strongly suspected for binarity. Using the first collected data for a limited sample we proved this method to be about 95\%-efficient, confirming SB2 status for 17 out of 18 analyzed Cepheids in the Large Magellanic Cloud (see P21 and \citealt{cepgiant1pta_2022}). Currently, the whole sample for this galaxy consists of 47 objects. It is interesting to note, that two of the Cepheid candidates for SB2 systems selected in P21 are in fact double Cepheids. These systems were not analyzed there and will be presented here together with systems from the Milky Way and the Small Magellanic Cloud.

In this paper, we present 9 candidates for binary double (BIND) Cepheids and show the first results of our observing program, which prove their binarity. Note that as BIND Cepheids we treat only those binary systems composed of two Cepheids for which the anticorrelated orbital motion is confirmed either spectroscopically or astrometrically. In Section~\ref{sec:data} we describe the sample and present the photometric and spectroscopic data used in the study. In Section~\ref{sec:anal} we show the results from the analysis of the presented data, including the preliminary orbital solutions. In Section~\ref{sec:conclusions} we draw conclusions from these results and describe the prospects of the project.

\section{Data} \label{sec:data}

\subsection{Object selection} \label{sec:objects}

From the newest catalogs of classical Cepheids of the OGLE project \citep{Soszynski_2017AcA_OCVS_MC_Cep, Udalski_2018_GalCep_EclBin} we selected 9 classified as double Cepheids, i.e. objects where presence of two Cepheids was detected at the exactly same coordinates. These objects are our candidates for binary systems composed of two Cepheids. An alternative possibility is that these stars are just a superposition (a blend) of two unrelated stars. This is unlikely as even a slight difference in coordinates would result in a variable shift in the photo-center at different pulsation phases. Nevertheless, to undoubtedly confirm that the two Cepheids that form a double Cepheid are gravitationally bound, spectroscopic confirmation of their anticorrelated orbital motion is necessary. Trying to obtain such a confirmation we aim to have, together with the published OGLE-LMC-CEP-1718 system, a relatively large number of ten BIND Cepheids for follow-up studies.

In our sample two objects are located in the Milky Way (MW), five in the Small Magellanic Cloud (SMC) and two in the Large Magellanic Cloud (LMC), thus covering a significant range of metallicities. None of them was studied before but it is reassuring that the LMC targets were identified in P21 having similar properties to other overbright Cepheids, for which evidence of their binarity was shown \citep[see also][]{cep1347_ApJL_2022}. We note here that the Magellanic Clouds Cepheids were listed by \citet{Szabados_2012_BinCep_MC} as known binaries but this was based only on them being double, with no direct evidence of their binarity provided. As all the objects come from the OGLE catalog, in the rest of the text we will omit the "OGLE-" prefix in their IDs.

The OGLE catalog provides individual pulsation modes of the components, which were obtained through the Fourier decomposition technique \citep{1981ApJ...248..291S}. Using these modes we can see that our targets represent all combinations of fundamental (F) and first-overtone (1O) Cepheids, i.e. pairs of F+F, F+1O and 1O+1O Cepheids.  
Their periods range from 1.1 to 4.6 days, with period ratios between the components ($P_2/P_1$) from 0.56 to 0.99. For comparison, the only known binary double Cepheid, LMC-CEP-1718, is composed of two 1O Cepheids, with periods of about 2 and 2.5 days ($P_2/P_1 \sim 0.79$). 
The monitoring of our proposed sample may thus not only lead to a 10-fold increase in the number of double binary Cepheids with known mass and radius ratios, but will also cover a considerably wide parameter space for Cepheids pulsating in different modes and located in environments with different typical metallicities. Such a sample will be of extreme value for testing pulsation and evolution theory models.

Basic parameters for double Cepheids from our sample are summarized in Table~\ref{tab:basic}. We name components with higher fundamental or fundamentalized (in case of first-overtone Cepheids) period as component A, and the other as component B. In P21 we provided for the LMC Cepheids a formula for fundamentalization, which on average maintains "luminosity", i.e. after the fundamentalization of periods we wanted the 1O Cepheids to lie on average on the same period-luminosity relation (period-Wesenheit index in this case) with F-mode Cepheids. For the current purpose, we obtained in the same way a formula for the Cepheids in the SMC. Both relations, that were applied to the LMC and SMC objects, are given below:

\begin{equation}
\begin{split}
(LMC)\quad P_F =  P_{1O} * (1.418 + 0.115 \log P_{1O}), \\
(SMC)\quad P_F =  P_{1O} * (1.433 + 0.096 \log P_{1O}).
\end{split}
\end{equation}

\noindent As the difference between these two relations, in terms of resulting ratios $P_F/P_{1O}$, is insignificant and our aim is mostly to make a comparison between the stars easier, we decided to use the LMC relation also for one 1O+1O MW Cepheid (note that period ratios between the same-mode Cepheids are very insensitive to the used transformation). It is worth to mention here that in the typical approach for fundamentalization the relation for $P_F/P_{1O}$ comes from period ratios in double-mode Cepheids (see, e.g., \citealt{Feast_1997_fundper,Kovtyukh_2016_fundper_feh}) with no constraint for the luminosity. We used the approach described in P21 principally because it serves better for comparing luminosities.

As components of binary systems should have the same age, and Cepheids are mostly found in a very specific stage of evolution (the blue loop), they should also have a similar mass and not very different radii (in a range allowed by the width of the instability strip and extent of the blue loop). For example, LMC-CEP-1718 ($P_2^F/P_1^F=0.786$) has the mass ratio of 0.98 and the radius ratio of 0.84 \citep{allcep_pilecki_2018}.
As pulsation period depends principally on the mass and radius, a presence of Cepheids with very different periods ($P_2^F/P_1^F \sim 0.6$) in the same system would be very intriguing, possibly meaning   1) a combination of first-crossing (still on the subgiant branch) and typical (on the blue loop) Cepheids, or   2) a stellar merger event in the system past evolution, which can lead to Cepheids of different masses having the same age. These options seem to have rather low probability of occurrence, but both are possible scenarios for the double Cepheids with low period ratios listed in Table~\ref{tab:basic} (once their binarity is confirmed)  and to discriminate between them a spectroscopic mass ratio will be necessary.

\begin{table*}
\centering
\caption{Basic data for known double Cepheids \label{tab:basic}}
\begin{tabular}{cc|cc|cc|ccc|c}
\hline
             &       & \multicolumn2c{component A}   & \multicolumn2c{component B}   &               &         &         &      \\
OGLE ID      & modes & $P_1$ [days] & $P_1^F$ [days] & $P_2$ [days] & $P_2^F$ [days] & $P_2^F/P_1^F$ & V [mag] & I [mag] & refs \\
\hline
BLG-CEP-067  & 1O+1O & 2.610721 & 3.827 & 1.692381 & 2.444 & 0.639 & 16.33 & 14.51 & (3) \\
GD-CEP-0291  & F+F   & 3.667693 & 3.668 & 3.398977 & 3.399 & 0.927 & 14.65 & 12.70 & (3) \\
LMC-CEP-0571 & F+1O  & 3.079937 & 3.080 & 2.100885 & 3.057 & 0.992 & 15.70 & 14.86 & (1,2) \\
LMC-CEP-0835 & F+F   & 4.562781 & 4.563 & 2.750956 & 2.751 & 0.603 & 15.25 & 14.46 & (1,2) \\
LMC-CEP-1718 & 1O+1O & 2.480909 & 3.649 & 1.963683 & 2.869 & 0.786 & 15.19 & 14.51 & (1,2,4) \\
SMC-CEP-1526 & F+F   & 1.804311 & 1.804 & 1.290234 & 1.290 & 0.715 & 16.83 & 16.16 & (2) \\
SMC-CEP-2699 & 1O+F  & 2.562225 & 3.772 & 2.117341 & 2.117 & 0.561 & 16.06 & 15.38 & (2) \\
SMC-CEP-2893 & F+F   & 1.321549 & 1.321 & 1.135859 & 1.136 & 0.860 & 16.93 & 16.38 & (2) \\
SMC-CEP-3115 & F+F   & 1.251945 & 1.252 & 1.159784 & 1.160 & 0.926 & 16.66 & 16.18 & (2) \\
SMC-CEP-3674 & F+1O  & 2.896089 & 2.896 & 1.827785 & 2.665 & 0.920 & 15.79 & 15.13 & (2) \\
\hline
\end{tabular}

\tablefoot{Period ($P_i$) and fundamentalized period ($P_i^F$) are given for each component (i=1,2) of a given double Cepheid. LMC-CEP-1718 is the only one already spectroscopically confirmed. References are given in the last column. }
\tablebib{(1) - \citet{Alcock_1995_double_cepheids}, (2) - \citet{Soszynski_2017AcA_OCVS_MC_Cep}, (3) - \citet{Udalski_2018_GalCep_EclBin}, (4) - \citet{allcep_pilecki_2018}. }
\end{table*}

\subsection{Photometry}

The majority of the photometric data used to in this work come from the OGLE project \citep{Soszynski_2017AcA_OCVS_MC_Cep, Udalski_2018_GalCep_EclBin}. Specifically, we used both the V and I-band light curves from the catalog of the OGLE-3 and OGLE-4 phases. I-band light curves have on average 1280 points in total, while V-band light curves have only about 119 points. Average V and I-band magnitudes are given in Table~\ref{tab:basic}. In the periodicity analysis (Section~\ref{sec:period}) only the I-band data was used but whenever possible (i.e., for the LMC objects) the photometry was extended with the R-band data\footnote{downloaded from http://macho.anu.edu.au} (600 points on average) from the MACHO project \citep{macho2002alcock}.

\subsection{Spectroscopy} \label{sec:spec}

Spectroscopic monitoring of our sample started in October, 2020, with the exception of BLG-CEP-067 for which the first spectrum was obtained almost a year later, in September, 2021. The observations were performed using three very efficient instruments mounted on telescopes located in three distinct observatories in Chile, and took place until January, 2024.
Most of the acquired spectra were obtained with the MIKE spectrograph mounted on the 6.5-m Magellan Clay telescope at the Las Campanas Observatory. We have also obtained spectra (in service mode) with the UVES spectrograph on the 8.2-m VLT at ESO Paranal Observatory. The three brightest Cepheids ($V \leq 15.7$ mag) were also observed with the HARPS instrument mounted at the 3.6-m telescope at ESO La Silla Observatory.

For the analysis we used the reduced HARPS spectra downloaded from the ESO Archive\footnote{http://archive.eso.org}. The MIKE data were reduced using Daniel Kelson's pipeline available at the Carnegie Observatories Software Repository\footnote{http://code.obs.carnegiescience.edu}, and the UVES data were reduced using the ESO Reflex software and the official pipeline available at the ESO Science Software repository\footnote{http://www.eso.org/sci/software.html}.

For the identification of components in the spectra and the measurement of radial velocities (RVs) we used the Broadening Function (BF) technique \citep{Rucinski_1992AJ_BF,Rucinski_1999ASPC_BF_SVD} implemented in the RaveSpan code \citep{t2cep098apj2017}. This technique provides narrower profiles than the cross-correlation function method, which helps in the separation of components, increasing the chance of detecting a companion and improving the precision of the RV measurements.


\section{Analysis and results} \label{sec:anal}

\subsection{Period analysis} \label{sec:period}

We looked for periodicity of all selected objects using the I-band light curves from the OGLE project. We did not use the measurements in the V filter because of their insufficient number for this kind of analysis.
In the case of the LMC objects, additional analysis was done adding photometry from the MACHO project that precedes OGLE observations.

We analyzed all the data sets at the same time, constraining the phase coefficients to be the same for OGLE-3 and 4, while leaving the amplitude parameters unconstrained. This way we made sure that there is no phase shift between the data sets from different phases of the OGLE project. As the shape of the R-band light curve is quite similar to the I-band one, and we were not struggling for high accuracy here, we did the same for the MACHO data. Lack of discontinuities between the sets confirmed this approach to be correct.

As for double Cepheids the variabilities of two components are superimposed on each other, we first disentangled them (by iteratively subtracting one periodic variability and fitting the other) and treated each component separately in the analysis. The separated light curves are presented in Fig.~\ref{fig:lcs}. Individual Cepheids are identified by -A or -B after the ID, where A marks the Cepheid with the longer fundamental (or fundamentalized) period, which we assume to be more luminous of the two. In the analysis we assumed two models, one with a constant period ($P$) and another with a linear period change ($dP/dt$). The determined periods and the period changes are given in Table~\ref{tab:puls_ephem}. In the upper part we show the results for all Cepheids (using the OGLE data only), while in the lower part only for the LMC objects, using the combined OGLE+MACHO photometric data.

\begin{figure*}
    \centering
\includegraphics[height=3.5cm]{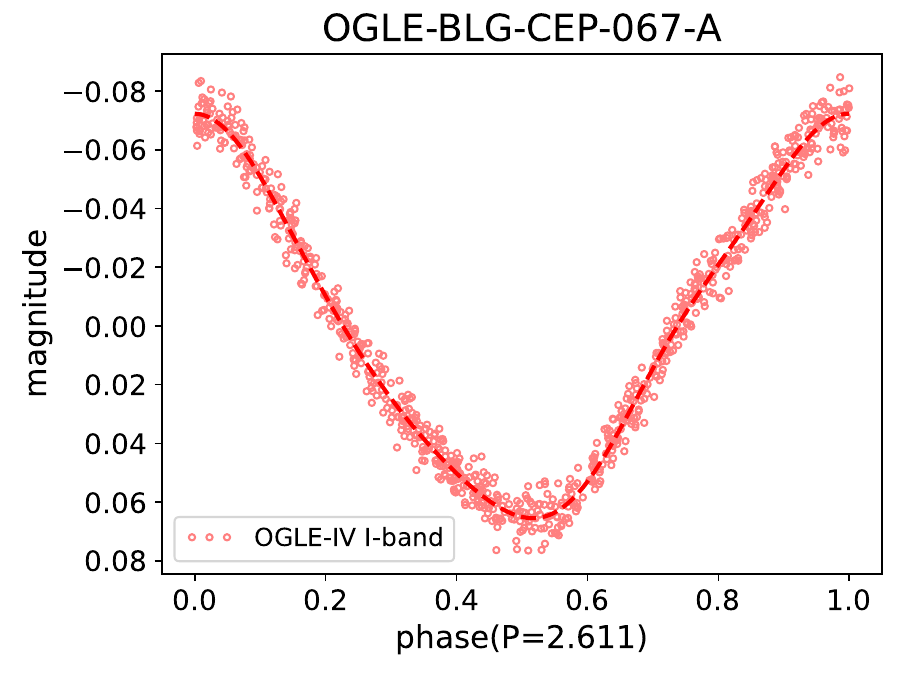}
\includegraphics[height=3.5cm]{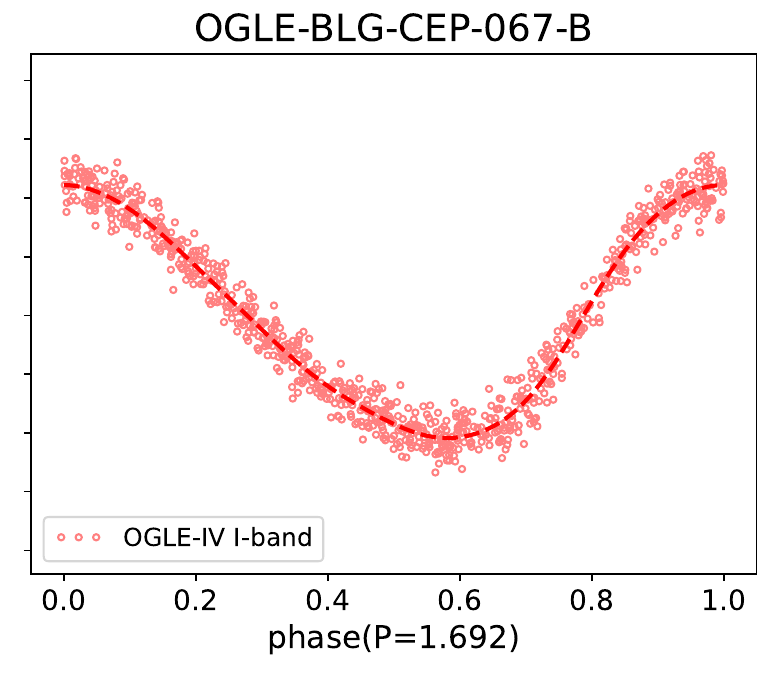}
\includegraphics[height=3.5cm]{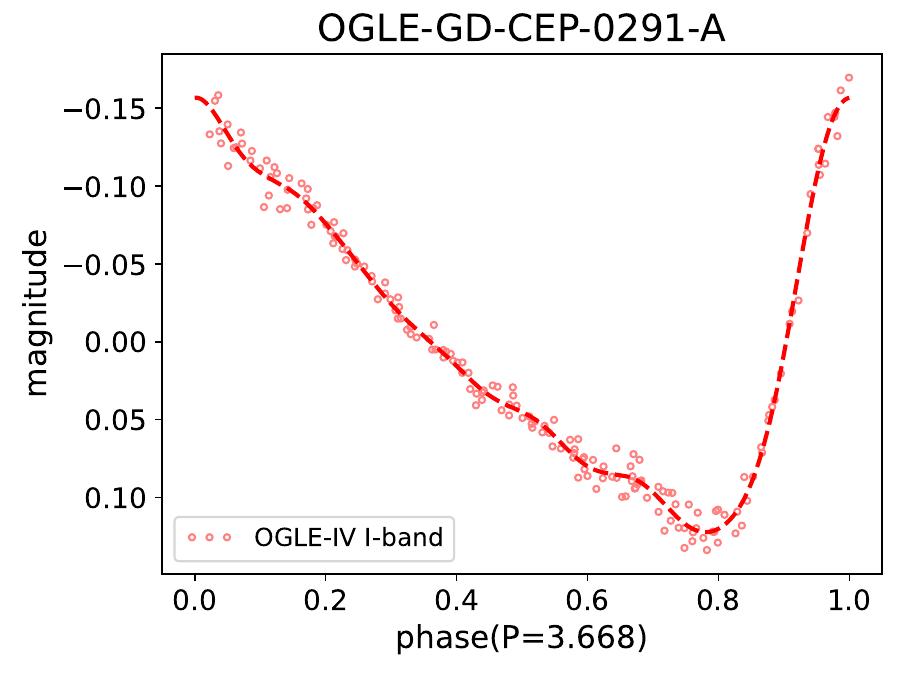}
\includegraphics[height=3.5cm]{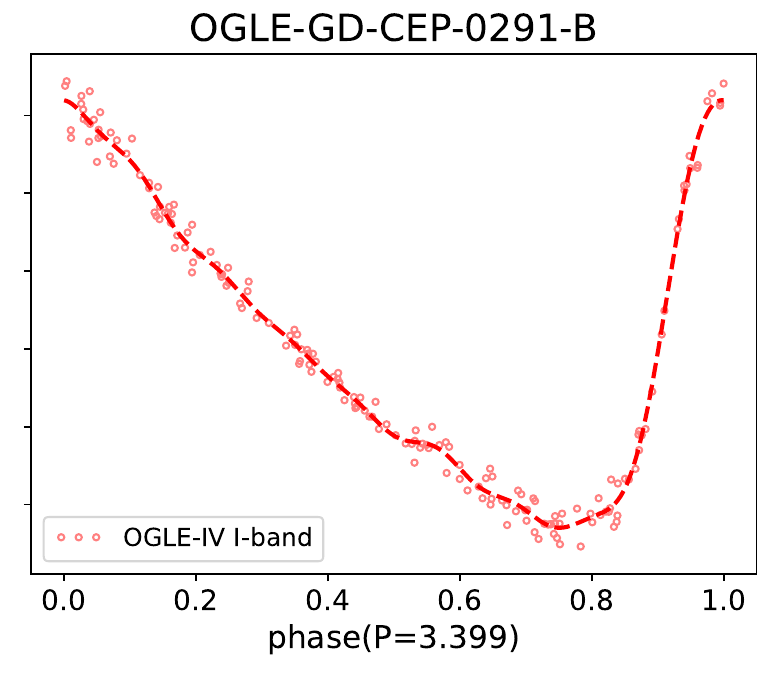}
\includegraphics[height=3.5cm]{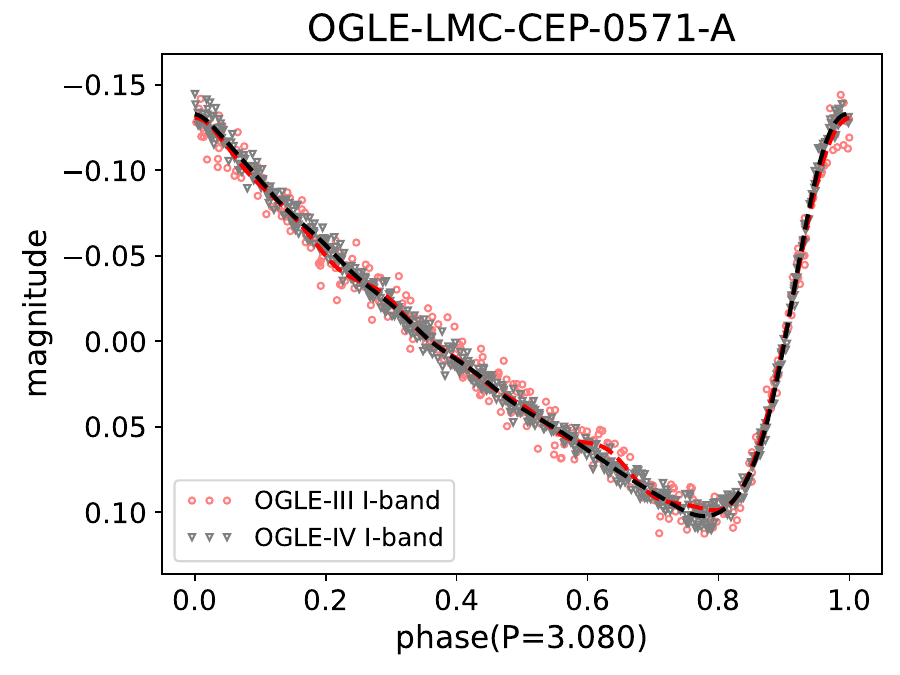}
\includegraphics[height=3.5cm]{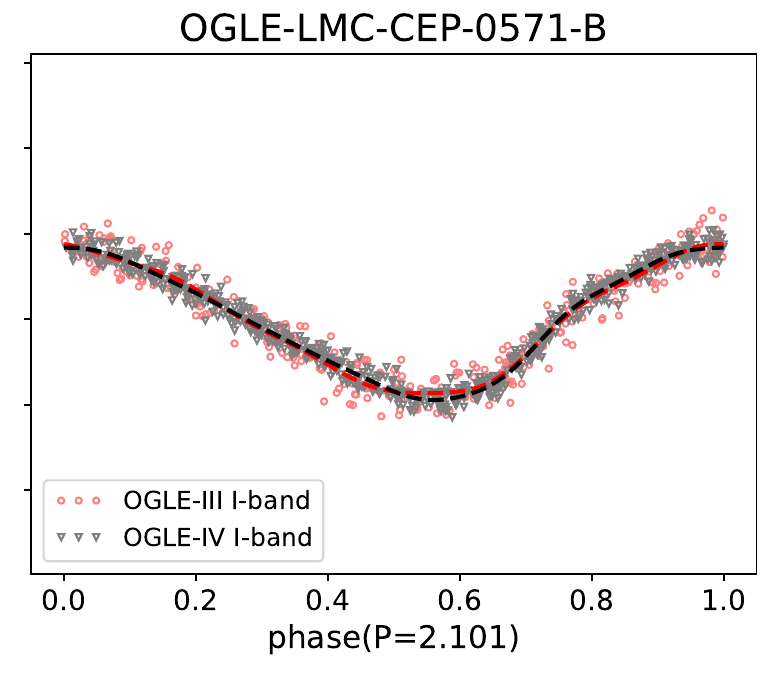}
\includegraphics[height=3.5cm]{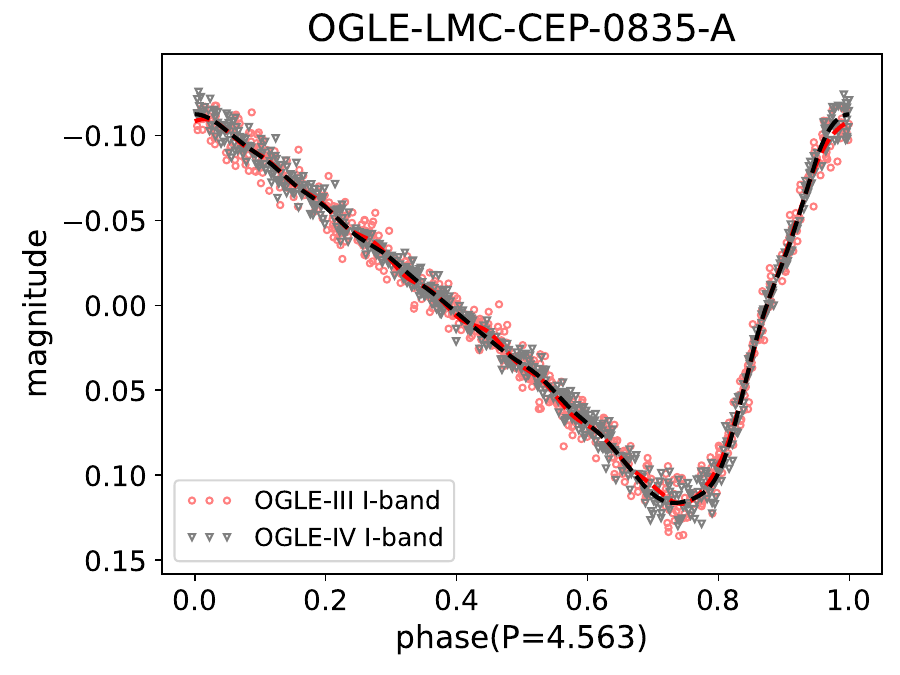}
\includegraphics[height=3.5cm]{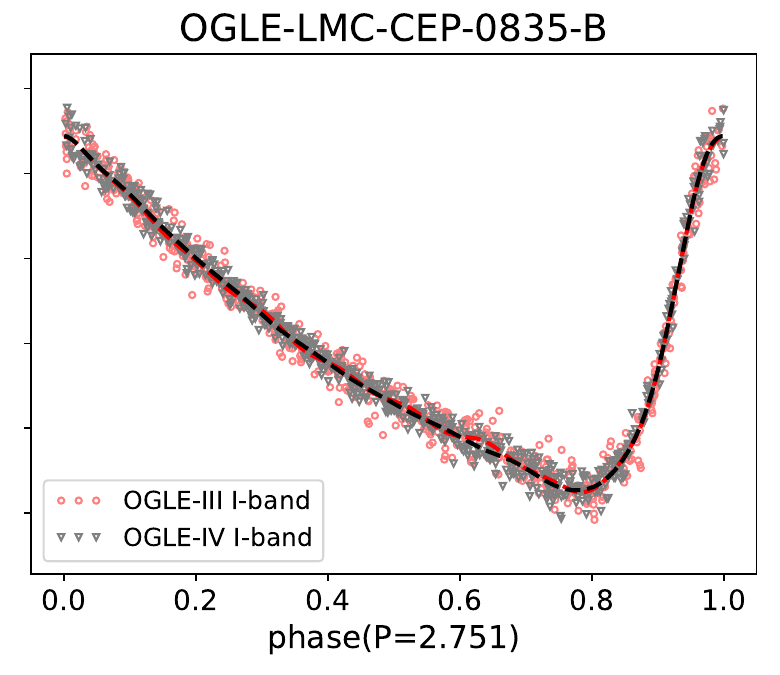}
\includegraphics[height=3.5cm]{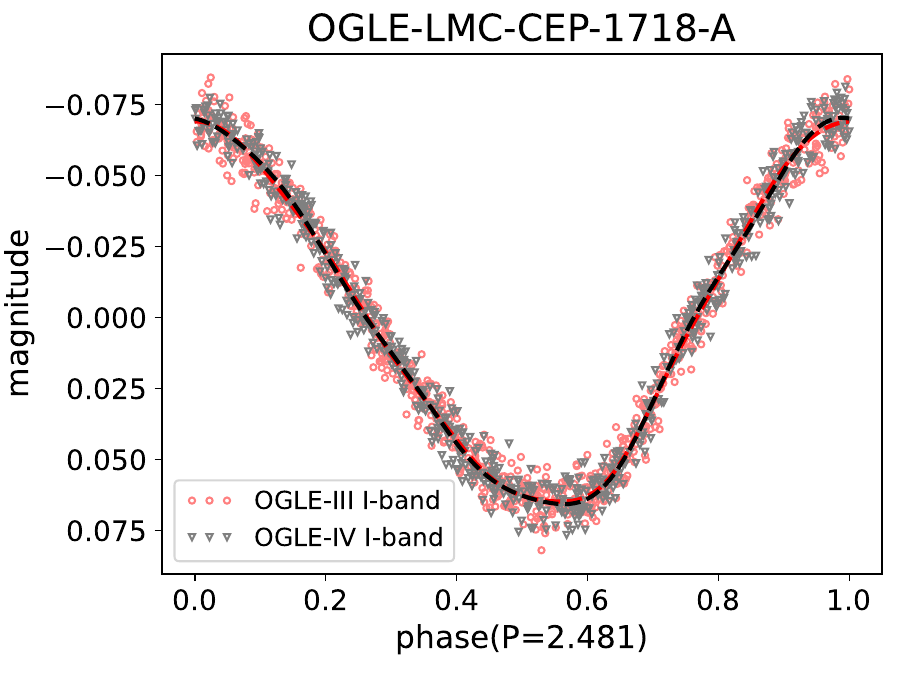}
\includegraphics[height=3.5cm]{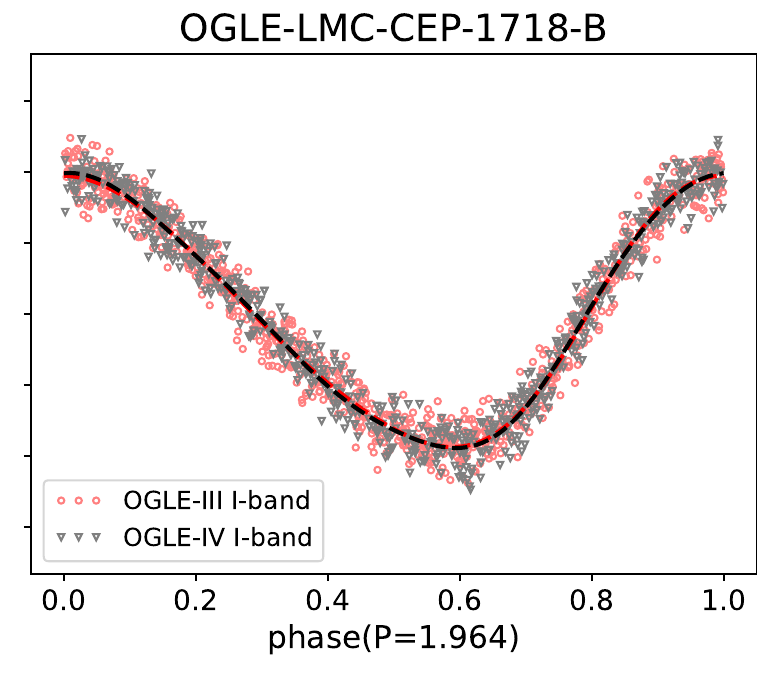}
\includegraphics[height=3.5cm]{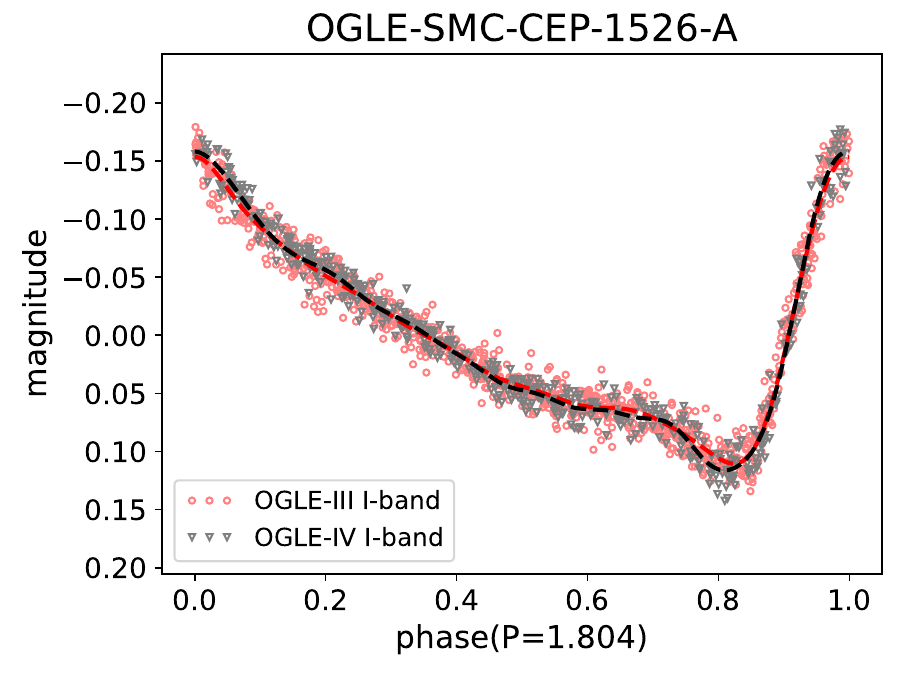}
\includegraphics[height=3.5cm]{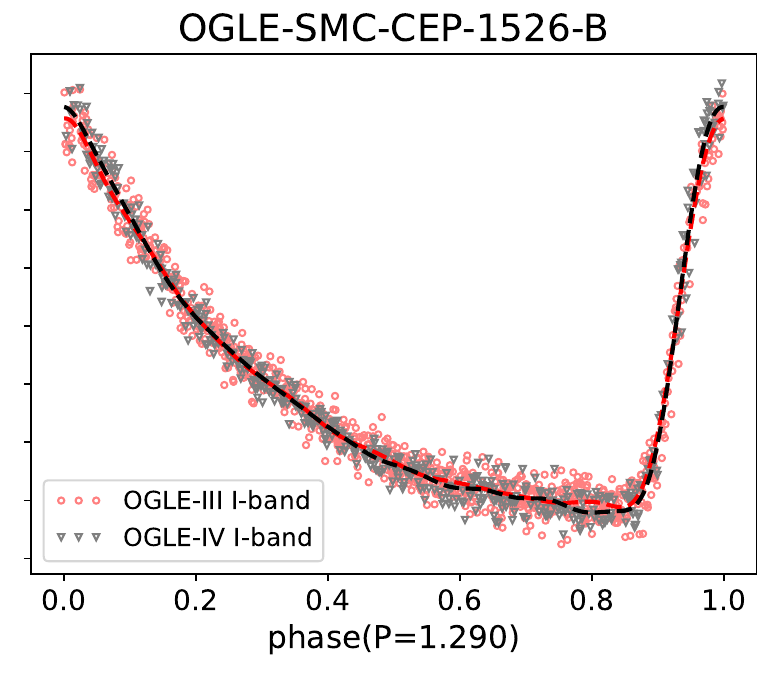}
\includegraphics[height=3.5cm]{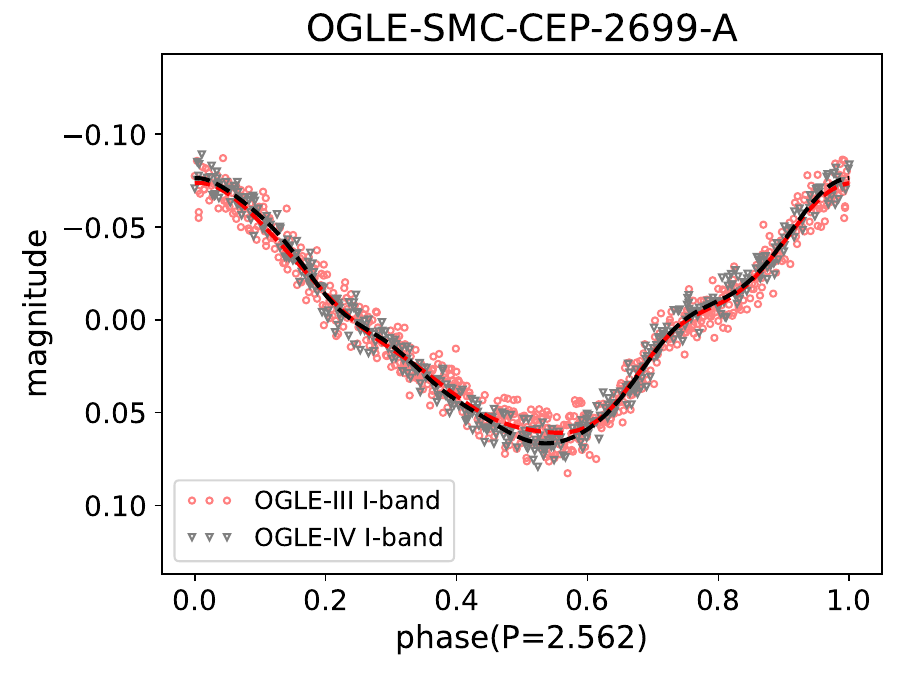}
\includegraphics[height=3.5cm]{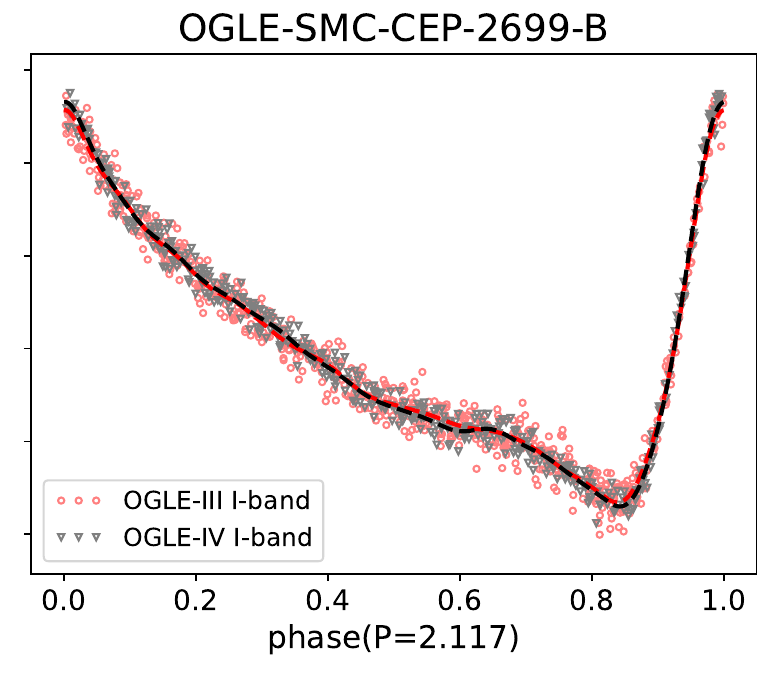}
\includegraphics[height=3.5cm]{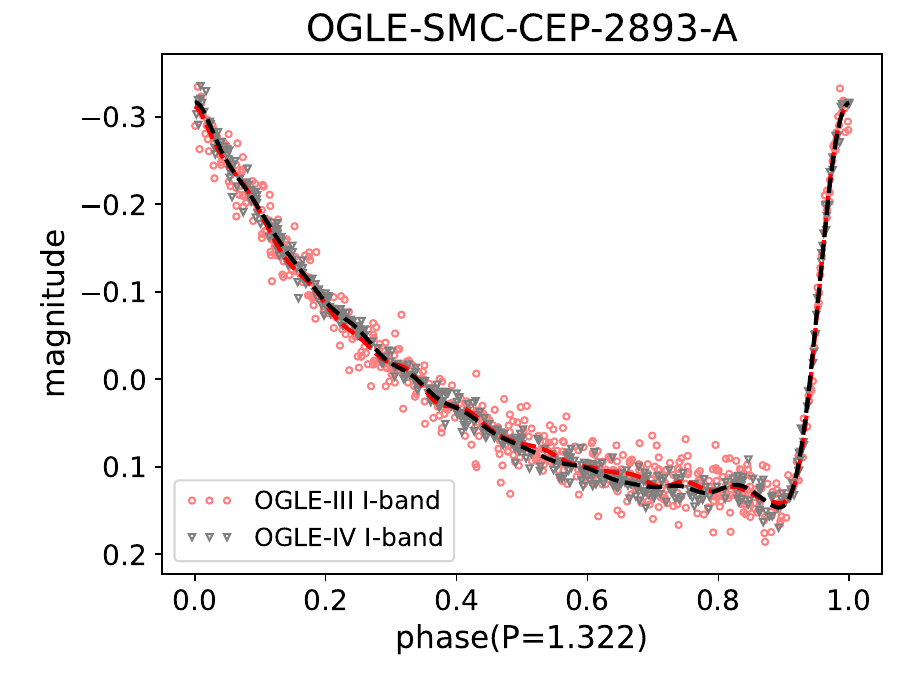}
\includegraphics[height=3.5cm]{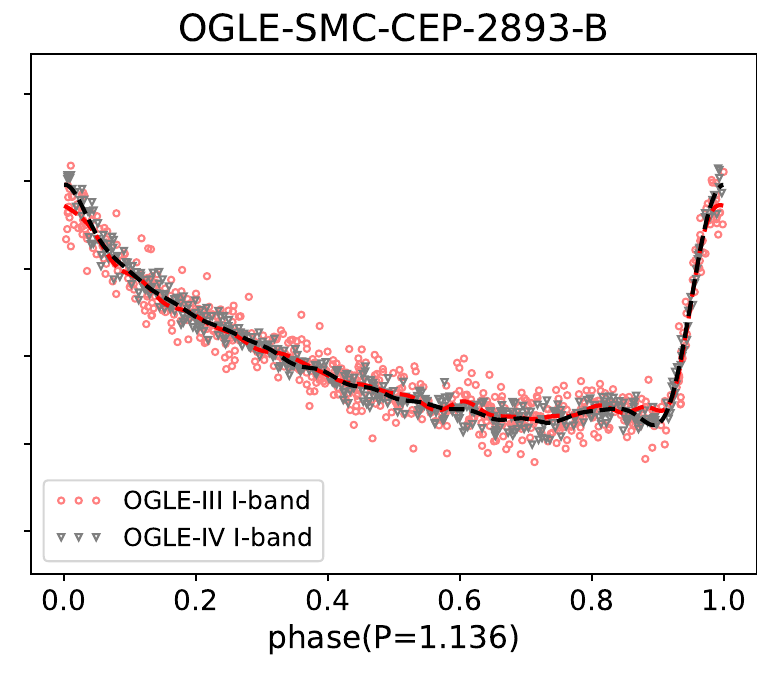}
\includegraphics[height=3.5cm]{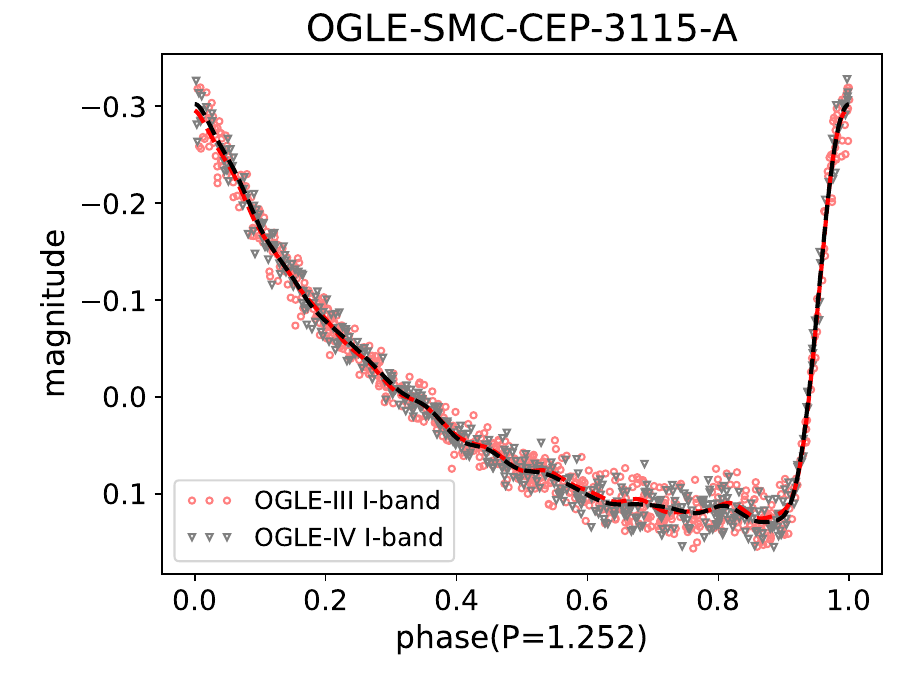}
\includegraphics[height=3.5cm]{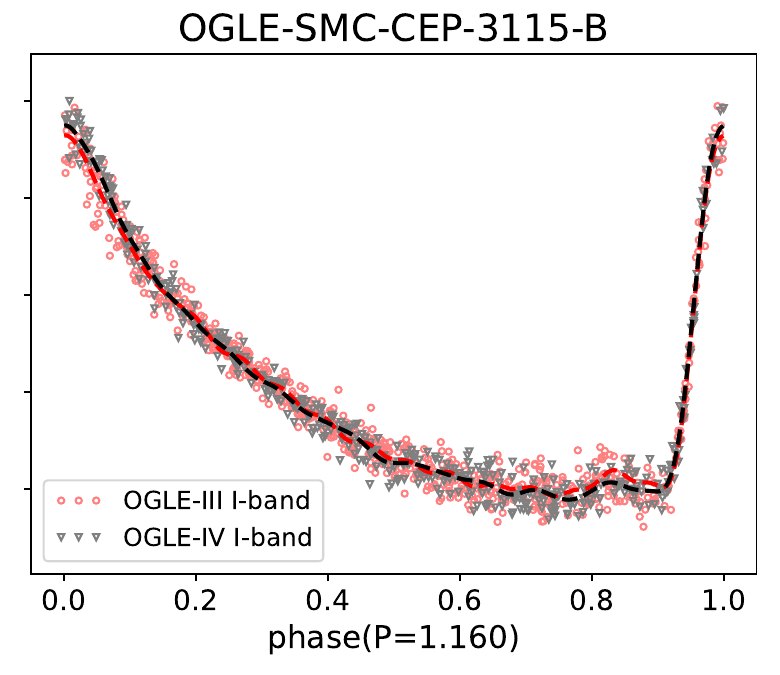}
\includegraphics[height=3.5cm]{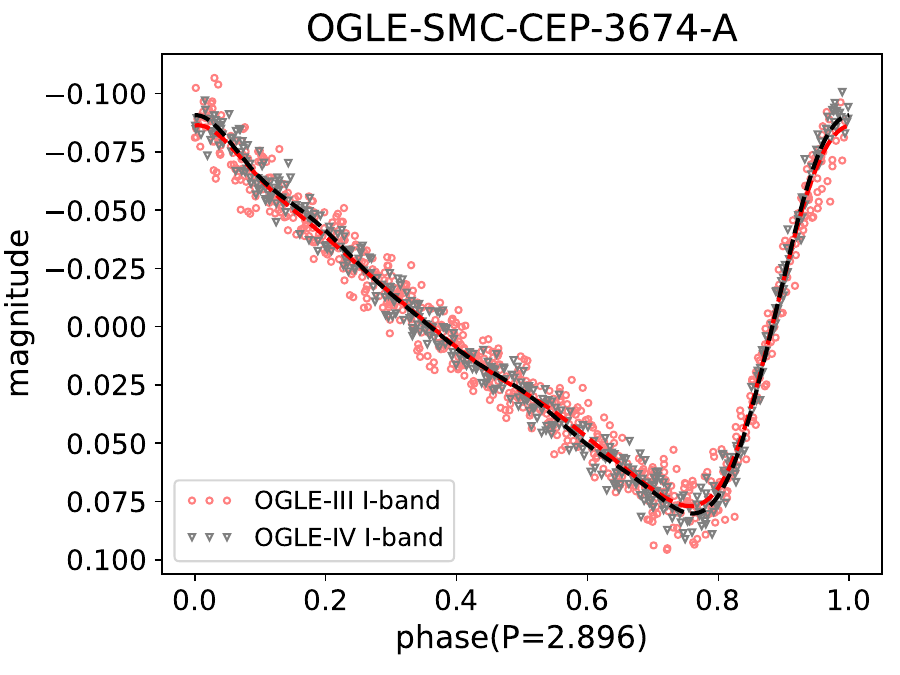}
\includegraphics[height=3.5cm]{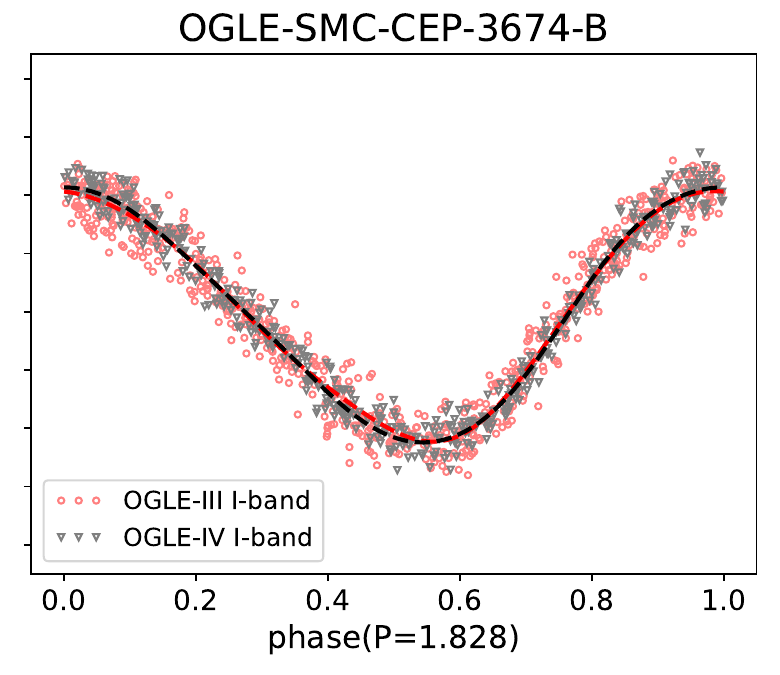}
    \caption{Pairs of OGLE I-band light curves for both components of ten double Cepheids phased with the ephemeris from Table~\ref{tab:puls_ephem} for linear period change. Span of Y-axis is the same for each pair for easier comparison of pulsation amplitudes. In general, components A (with longer periods) have higher amplitudes, with the exception of OGLE-SMC-CEP-1526 (F+F) and OGLE-SMC-CEP-2699 (1O+F).}
    \label{fig:lcs}
\end{figure*}

\begin{table*}
\centering
\caption{Pulsation ephemeris data for the components of double Cepheids  \label{tab:puls_ephem}}
\begin{tabular}{ccc|c|cc|ccc}
\hline
        &      &              & constant $P$  & \multicolumn2c{linear $dP/dt$} &             & O-C &       \\
OGLE ID & mode & $T_0$ [days] & period [days] & period[days] & $dP/dt$         & $A_I$ [mag] & $r$ & flags \\
\hline
BLG-CEP-067-A  & 1O & 6728.9156 & 2.610720(4)    & 2.610719(4)    &    2.1e$-$08 [2.0$\sigma$] & 0.138 & E+P & - \\
BLG-CEP-067-B  & 1O & 6728.1099 & 1.6923770(29)  & 1.6923836(27)  & $-$7.1e$-$08 [11$\sigma$]  & 0.086 & $-$0.60 & - \\
GD-CEP-0291-A  & F  & 7241.9130 & 3.667699(13)   & 3.667720(17)   & $-$7.8e$-$08 [2.0$\sigma$] & 0.279 & E+E & - \\
GD-CEP-0291-B  & F  & 7239.6049 & 3.398977(14)   & 3.398970(14)   &    5.6e$-$08 [1.5$\sigma$] & 0.275 & $-$0.42 & - \\
LMC-CEP-0571-A & F  & 5656.6539 & 3.0798876(21)  & 3.0798760(26)  & $-$1.7e$-$08 [7.0$\sigma$] & 0.230 & P+E & - \\
LMC-CEP-0571-B & 1O & 5633.7415 & 2.100785(4)    & 2.100815(4)    &    4.4e$-$08 [11$\sigma$]  & 0.088 & $-$0.94 & - \\
LMC-CEP-0835-A & F  & 5062.4994 & 4.562725(4)    & 4.562714(5)    & $-$2.5e$-$08 [3.5$\sigma$] & 0.226 & O+O? & LTTE \\
LMC-CEP-0835-B & F  & 4985.9788 & 2.7508653(16)  & 2.7508776(20)  &    2.3e$-$08 [9.0$\sigma$] & 0.210 & $-$0.83 & - \\
LMC-CEP-1718-A & 1O & 3923.5083 & 2.4809134(14)  & 2.4809078(14)  & $-$1.8e$-$08 [11$\sigma$]  & 0.134 & P+E & ECL \\
LMC-CEP-1718-B & 1O & 3914.7558 & 1.9636626(11)  & 1.9636615(12)  & $-$4.1e$-$09 [2.9$\sigma$] & 0.096 & 0.33 & - \\
SMC-CEP-1526-A & F  & 4038.4523 & 1.8043149(6)   & 1.8043146(6)   & $-$1.1e$-$09 [1.6$\sigma$] & 0.264 & O+O & LTTE \\
SMC-CEP-1526-B & F  & 3751.2473 & 1.29022591(26) & 1.29022616(25) &    1.6e$-$09 [5.1$\sigma$] & 0.334 & $-$0.97 & - \\
SMC-CEP-2699-A & 1O & 4942.2171 & 2.5622543(28)  & 2.5622527(29)  & $-$8.5e$-$09 [2.2$\sigma$] & 0.134 & E+P & LTTE \\
SMC-CEP-2699-B & F  & 4780.6972 & 2.1174356(12)  & 2.1174265(10)  & $-$3.9e$-$08 [29$\sigma$]  & 0.208 & 0.34 & - \\
SMC-CEP-2893-A & F  & 5326.5241 & 1.32155113(30) & 1.3215514(4)   &    3.7e$-$10 [0.8$\sigma$] & 0.453 & E+E & 1C \\
SMC-CEP-2893-B & F  & 5292.0427 & 1.1358590(4)   & 1.1358595(4)   &    9e$-$10 [1.9$\sigma$]   & 0.245 & 0.44 & - \\
SMC-CEP-3115-A & F  & 4558.5922 & 1.2519451(3)   & 1.25194563(29) &    4.3e$-$09 [10$\sigma$]  & 0.420 & O+O & LTTE \\
SMC-CEP-3115-B & F  & 4603.3492 & 1.15978962(26) & 1.15978938(25) & $-$4.2e$-$09 [11$\sigma$]  & 0.372 & $-$0.94 & - \\
SMC-CEP-3674-A & F  & 5009.1460 & 2.8960253(25)  & 2.8960260(28)  &    2.3e$-$09 [0.7$\sigma$] & 0.163 & O+O & LTTE \\
SMC-CEP-3674-B & 1O & 5008.9881 & 1.8277750(19)  & 1.8277743(20)  & $-$2.5e$-$09 [1.0$\sigma$] & 0.108 & $-$0.82 & - \\
\hline\hline    
LMC-CEP-0571-A & F  & 4929.8173 & 3.0799661(9)   & 3.0798934(19)  & $-$4e$-$08 [43$\sigma$]    & 0.228 & P+E & - \\
LMC-CEP-0571-B & 1O & 4820.7063 & 2.1007583(13)  & 2.1007901(28)  &    1.9e$-$08 [13$\sigma$]  & 0.088 & $-$0.76 & - \\
LMC-CEP-0835-A & F  & 4556.0275 & 4.5627231(27)  & 4.562725(4)    &    1.3e$-$09 [0.5$\sigma$] & 0.225 & E+O & LTTE \\
LMC-CEP-0835-B & F  & 4232.2570 & 2.7508940(10)  & 2.7508739(14)  & $-$1.8e$-$08 [21$\sigma$]  & 0.207 & $-$0.55 & - \\
LMC-CEP-1718-A & 1O & 3097.3722 & 2.4809333(10)  & 2.4809322(10)  & $-$2.5e$-$08 [23$\sigma$]  & 0.134 & P+E & ECL \\
LMC-CEP-1718-B & 1O & 3044.8611 & 1.9636670(8)   & 1.9636673(8)   & $-$5.1e$-$09 [5.5$\sigma$] & 0.095 & 0.81 & - \\
\hline
\end{tabular}

\tablefoot{Two ephemerides are presented, assuming either a constant period ($P$) or a linear period change ($dP/dt$). The same reference time ($T_0$, maximum brightness at I-band for a constant $P$) is used for both. Errors in the last digits are given in parentheses. For $dP/dt$ the significance in sigmas is given in brackets. In the penultimate column, a different information is given in rows for each component. O-C means the type of O-C variability for both Cepheids (E - erratic, P - parabolic, O - orbital motion), $r$ is the correlation parameter for these variabilities. Flags: 1C - at least 1 orbital cycle is covered, ECL - eclipsing system, LTTE - anticorrelated light travel-time effect detected. Results for the OGLE data only are given above the double horizontal line, while those using additional MACHO data are shown below.} 
\end{table*}


\subsection{O-C analysis} \label{sec:oc}

\begin{figure*}
    \centering
\includegraphics[width=0.245\linewidth]{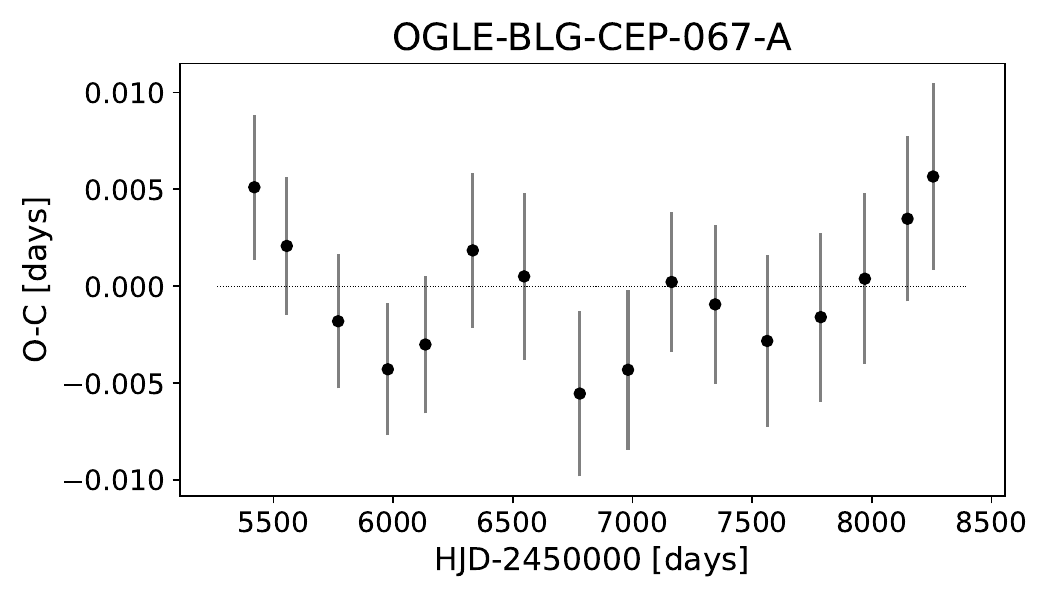}
\includegraphics[width=0.245\linewidth]{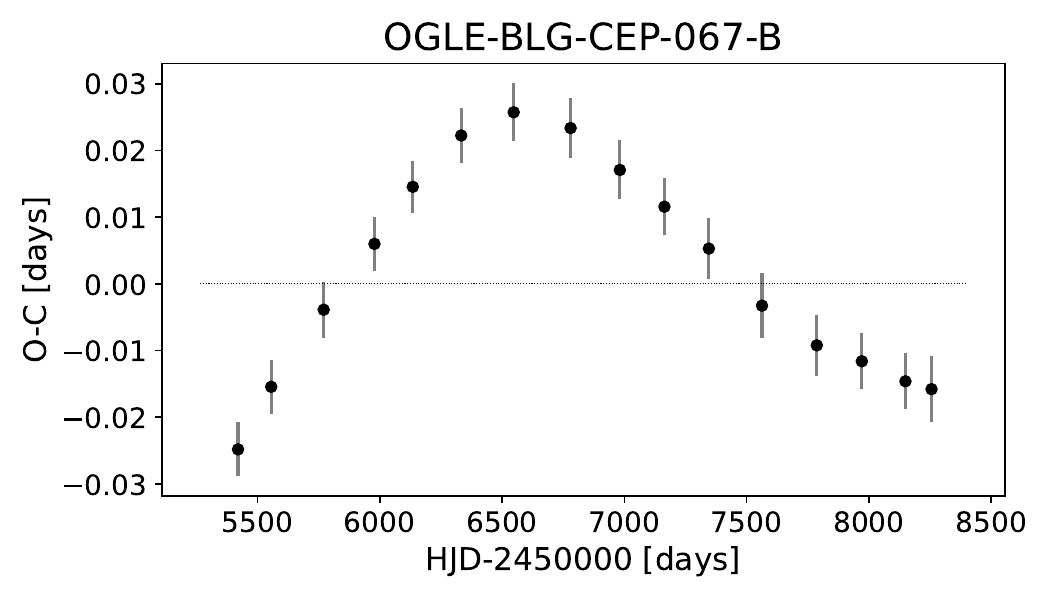}
\includegraphics[width=0.245\linewidth]{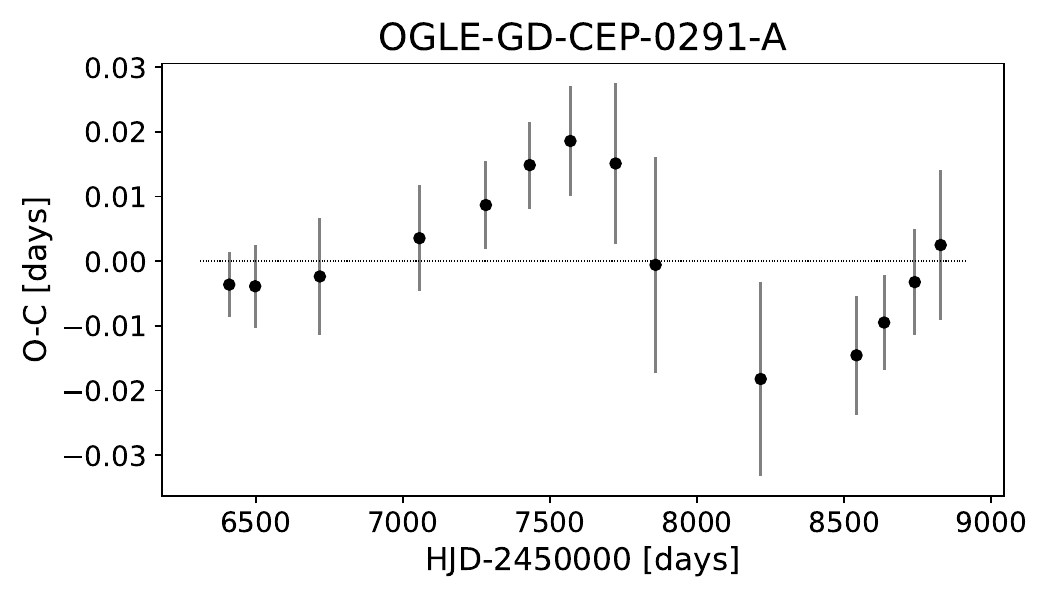}
\includegraphics[width=0.245\linewidth]{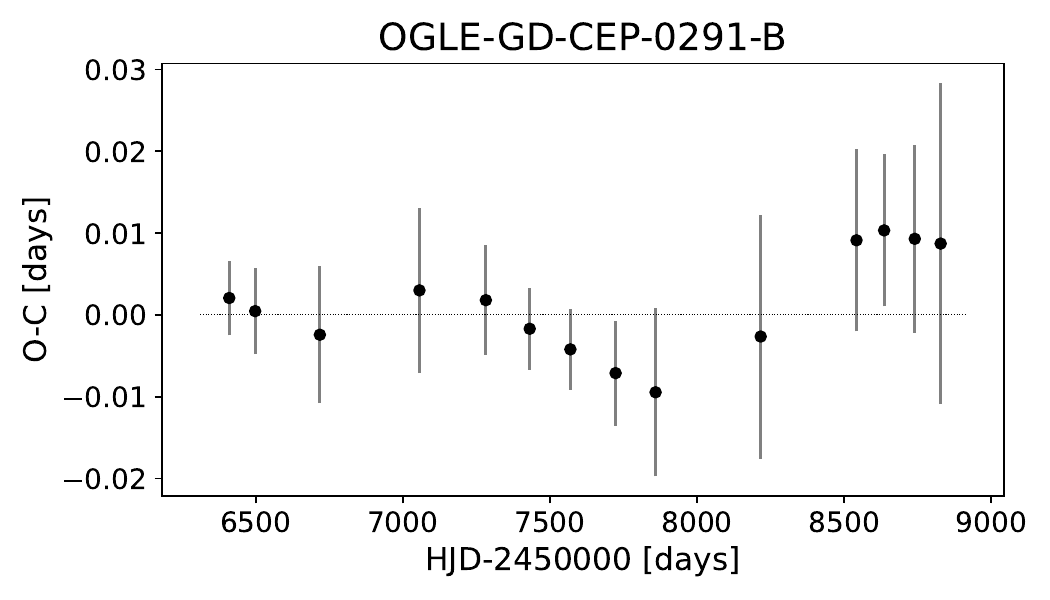}
\includegraphics[width=0.245\linewidth]{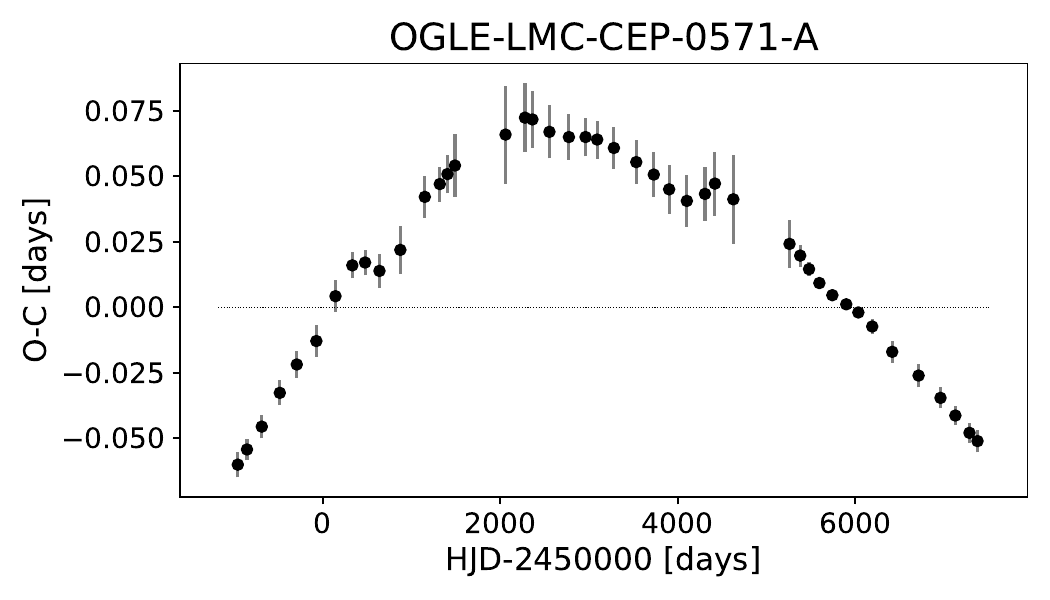}
\includegraphics[width=0.245\linewidth]{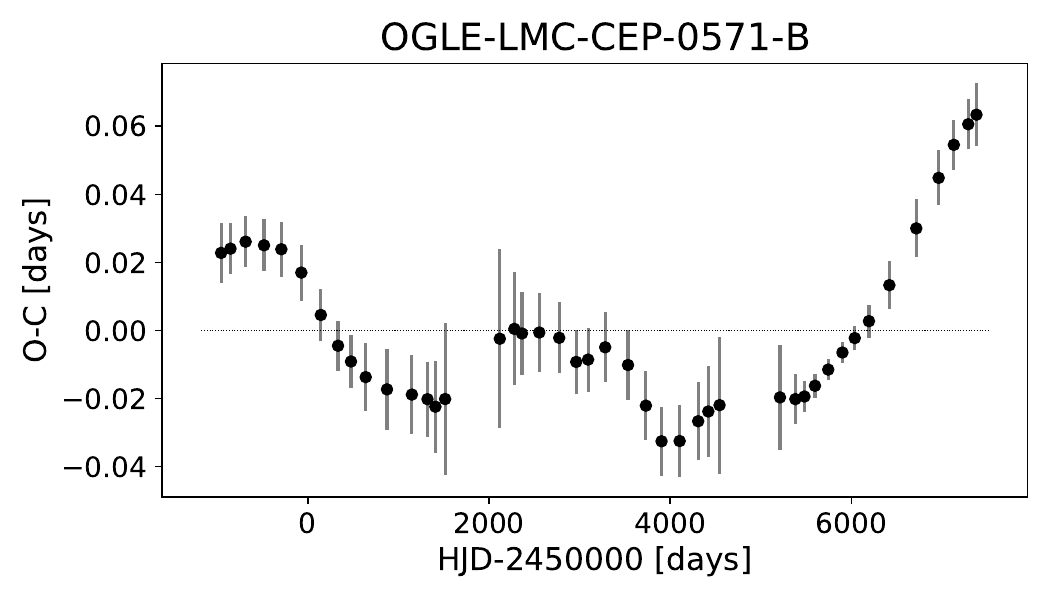}
\includegraphics[width=0.245\linewidth]{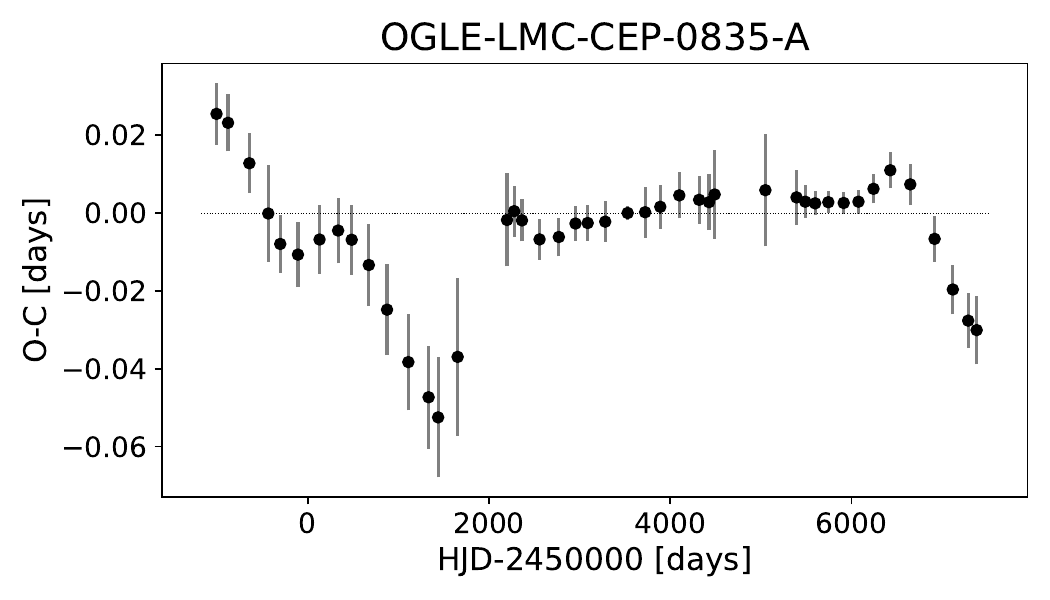}
\includegraphics[width=0.245\linewidth]{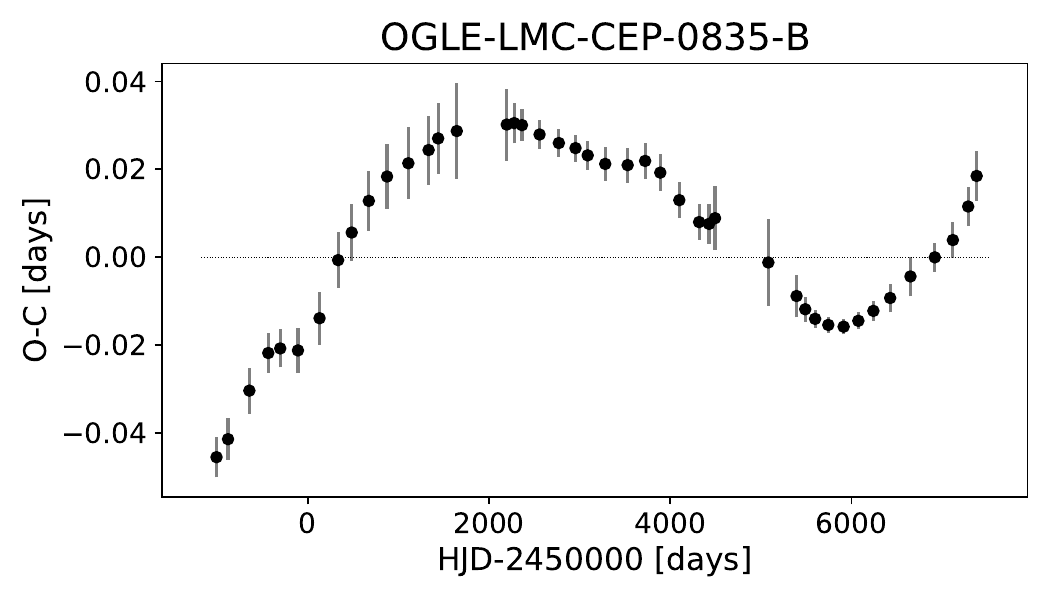}
\includegraphics[width=0.245\linewidth]{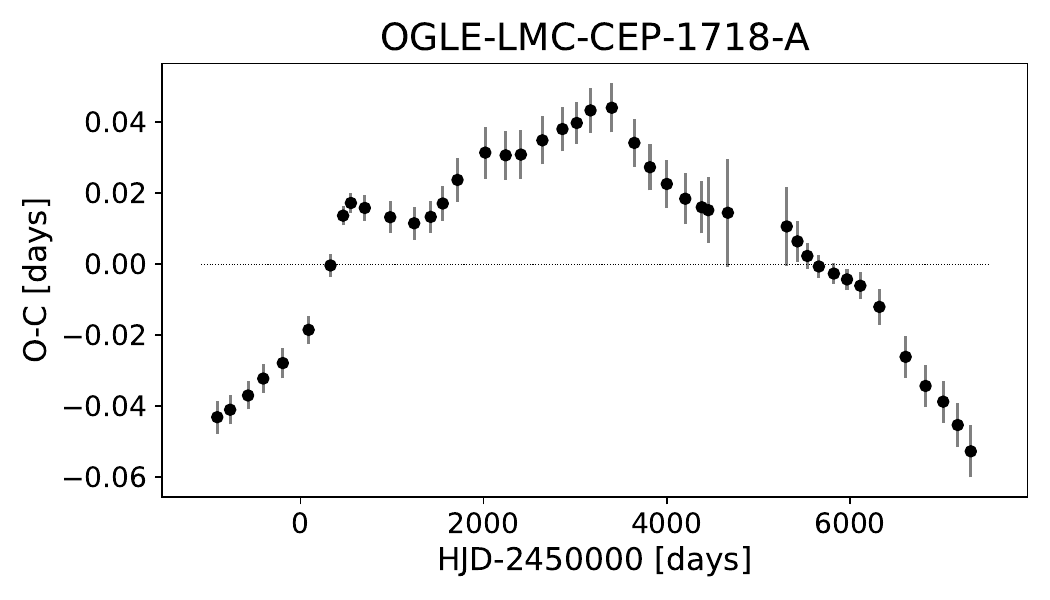}
\includegraphics[width=0.245\linewidth]{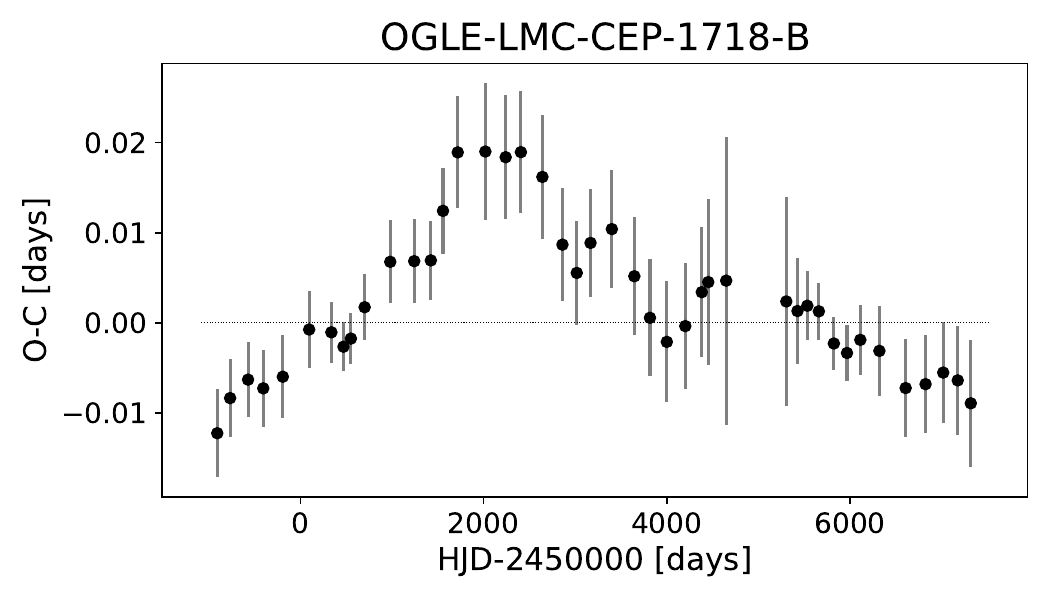}
\includegraphics[width=0.245\linewidth]{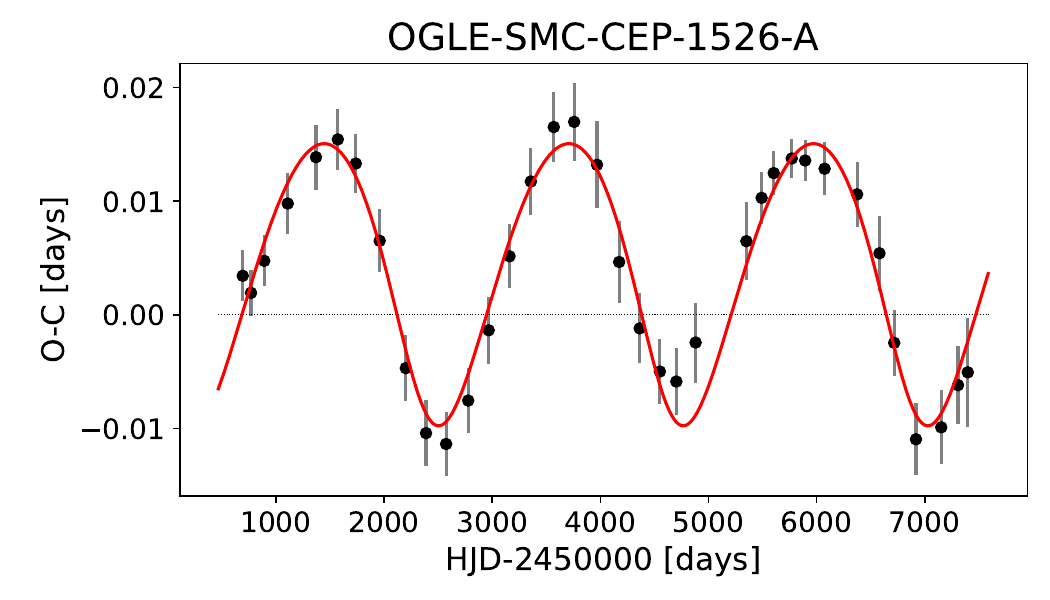}
\includegraphics[width=0.245\linewidth]{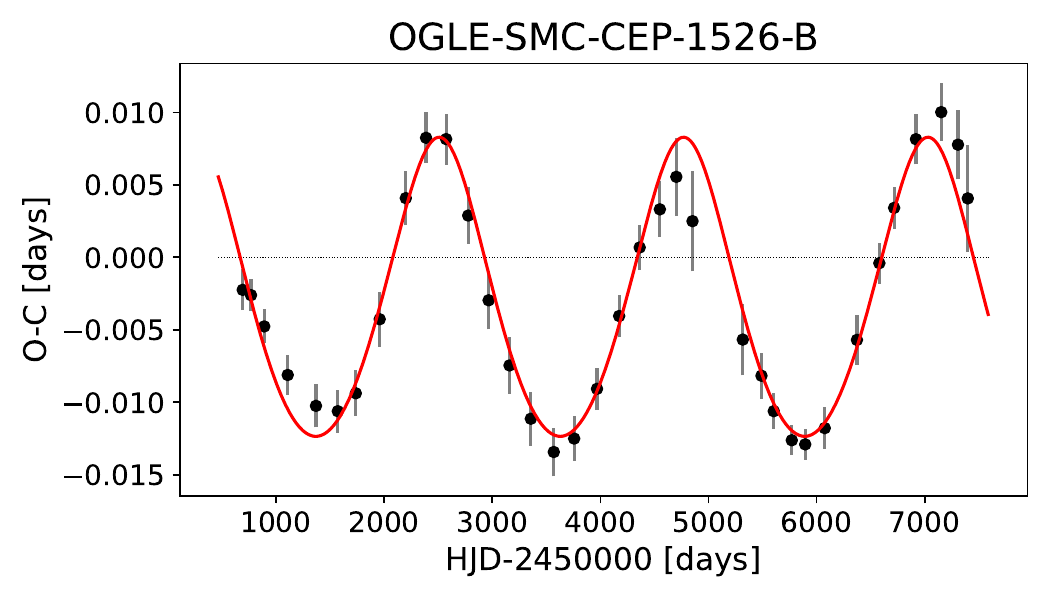}
\includegraphics[width=0.245\linewidth]{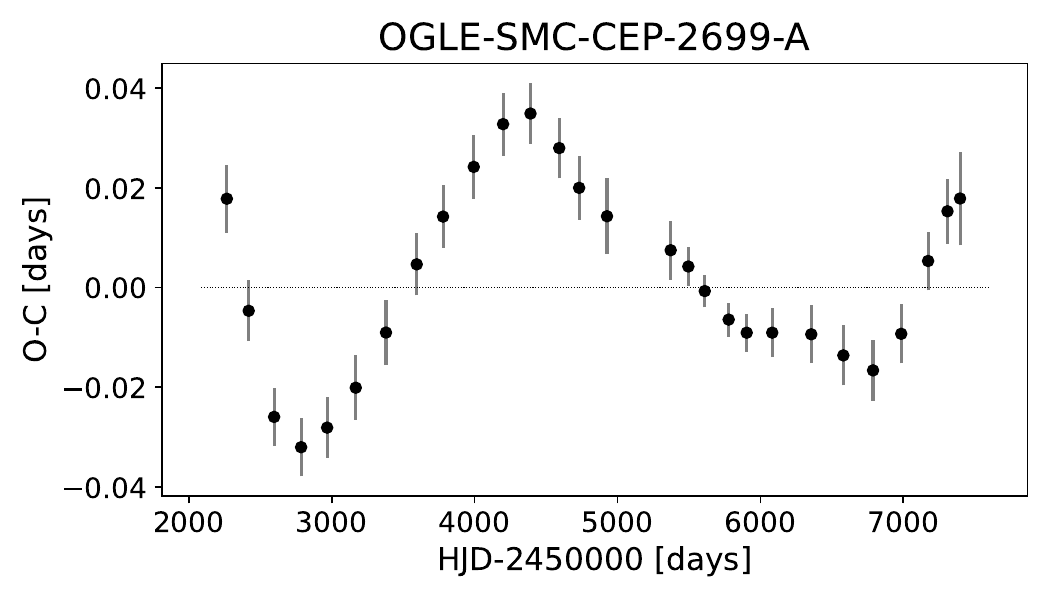}
\includegraphics[width=0.245\linewidth]{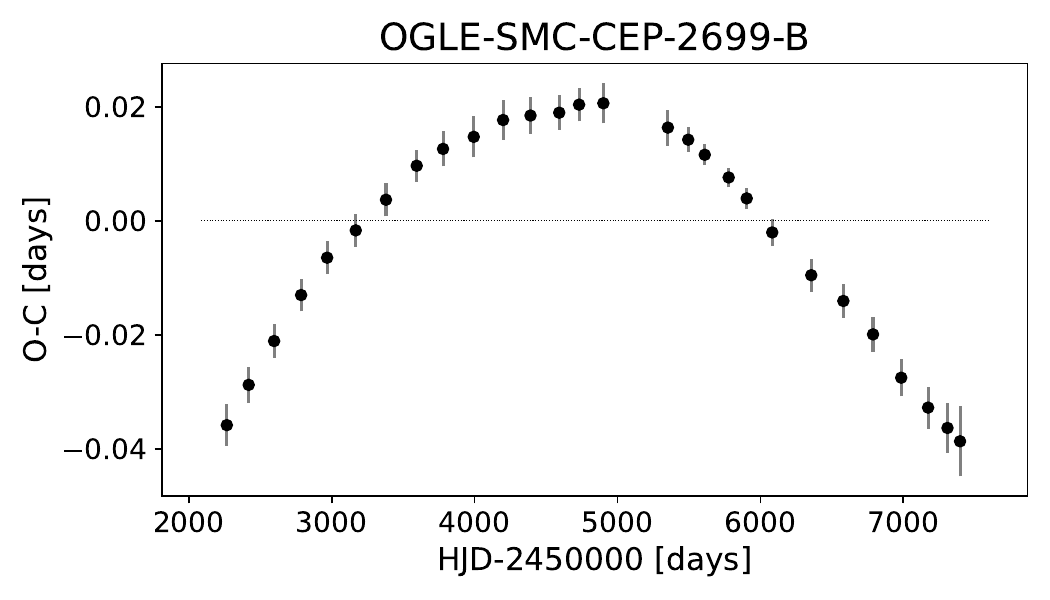}
\includegraphics[width=0.245\linewidth]{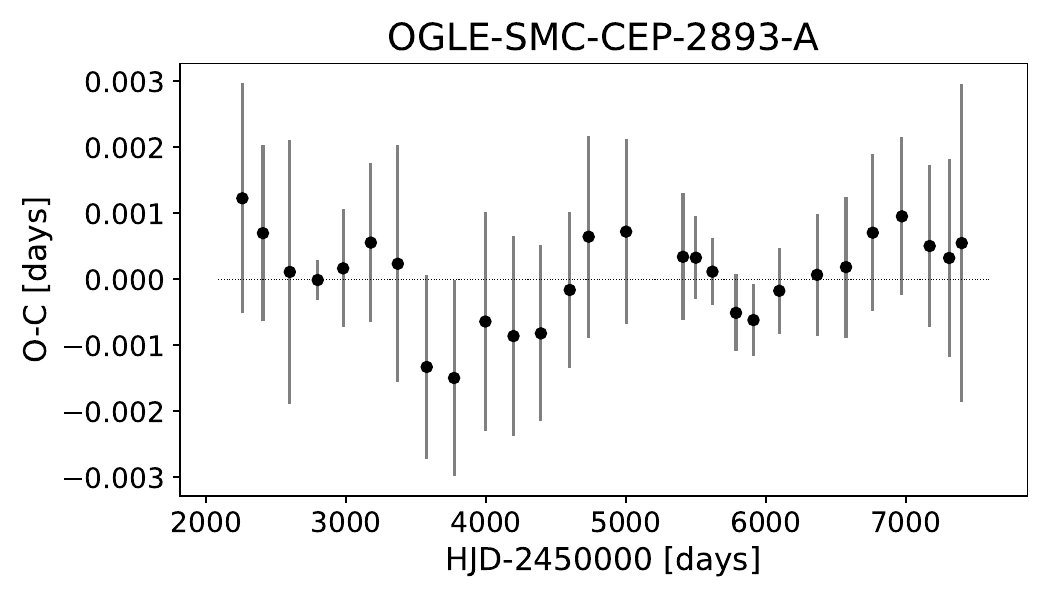}
\includegraphics[width=0.245\linewidth]{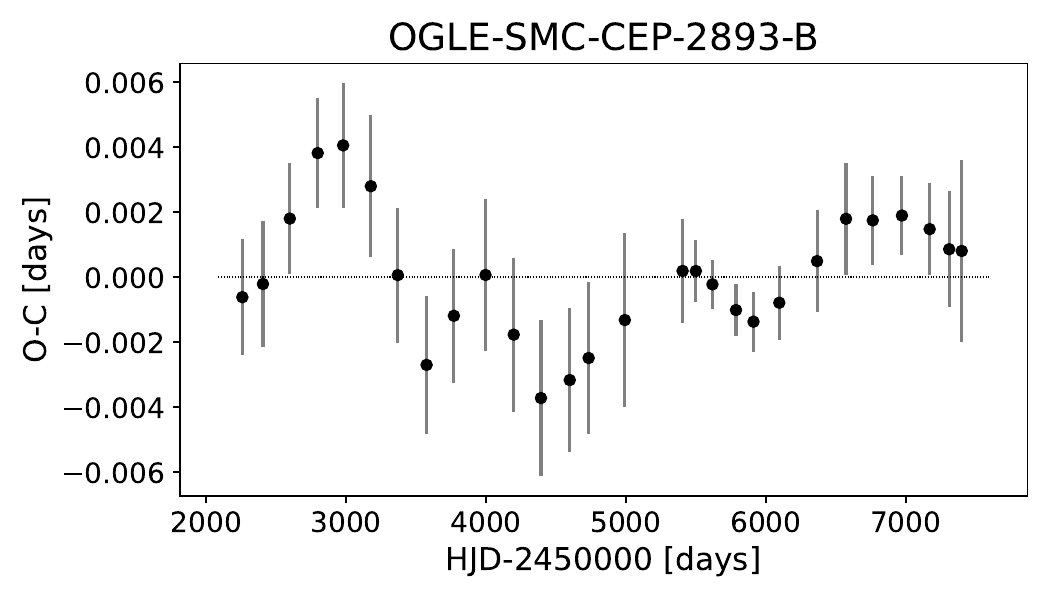}
\includegraphics[width=0.245\linewidth]{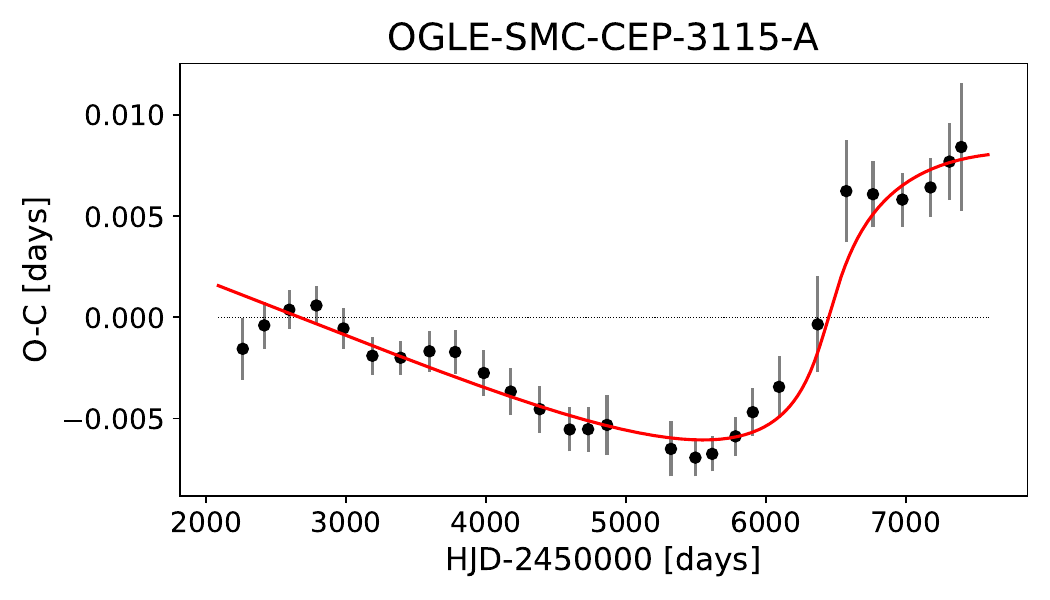}
\includegraphics[width=0.245\linewidth]{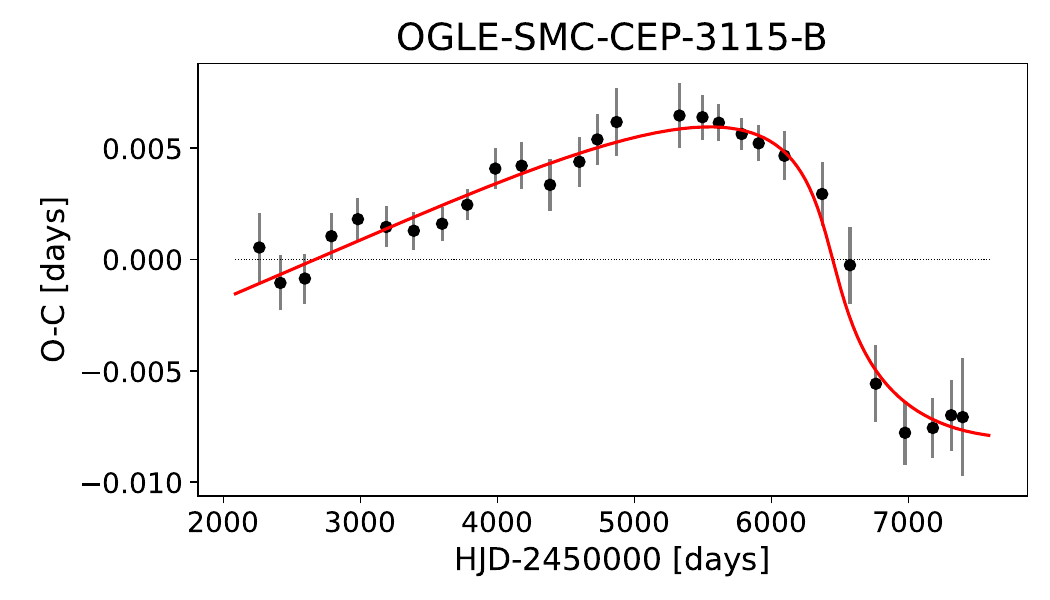}
\includegraphics[width=0.245\linewidth]{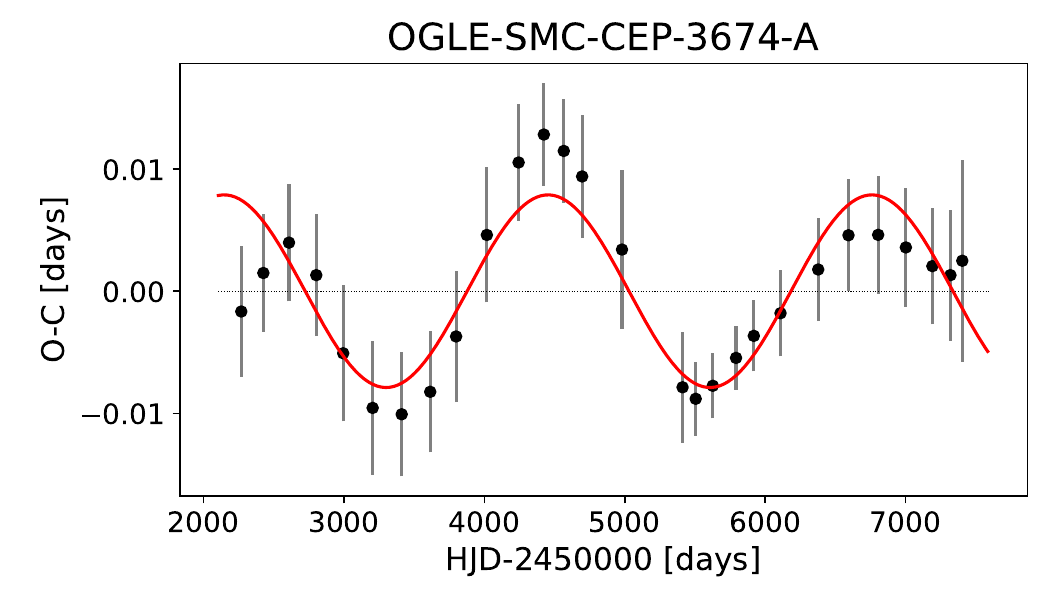}
\includegraphics[width=0.245\linewidth]{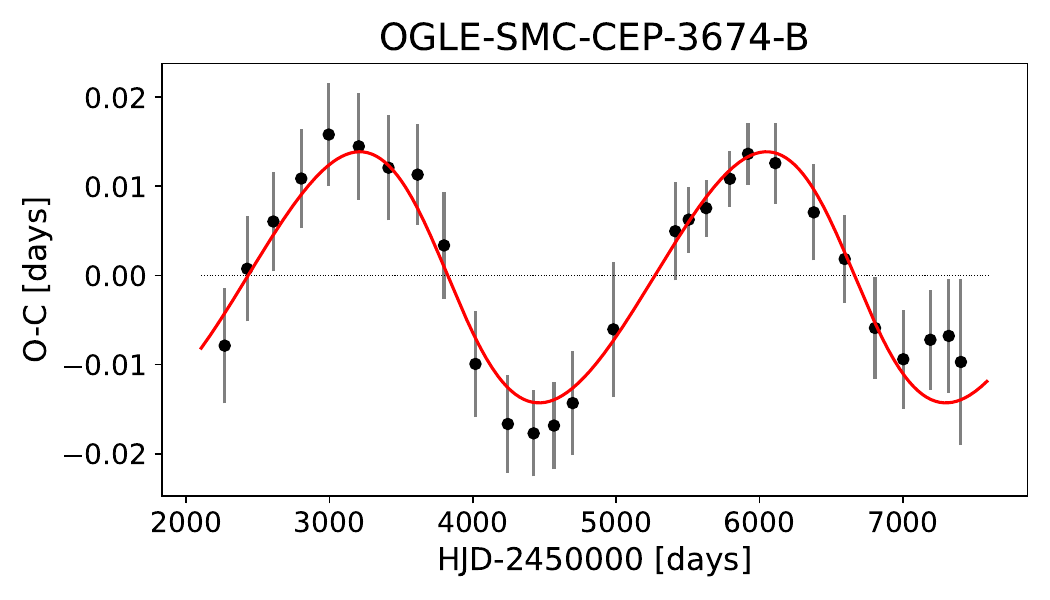}
    \caption{Pairs of O-C plots for both components of candidate binary double Cepheids. LTTE fit is shown with red line for systems with correlation parameter of O-C variations $r<-0.8$. Three systems (OGLE-SMC-CEP-1526,3115,3674) show very strong evidence for LTTE. Two other stars (OGLE-LMC-CEP-0835,0571) show anticorrelated phase shifts, but the conclusions are less clear (see text). We show a common orbital solution for them in Fig.~\ref{fig:ltte}.
    }
    \label{fig:ocs}
\end{figure*}

To test if these double Cepheids are gravitationaly bound for each disentangled pulsational variability we performed an O-C analysis, calculating instantaneous phase shifts along the collected photometric observational data. Then we looked for any sign of light travel-time effect (LTTE) due to the common binary motion of the components. The O-C diagrams and a binary motion fit (when applicable) are shown in Fig.~\ref{fig:ocs}.
As the amplitude of the phase variability in such analysis depends on the size of the projected orbit and the precision depends on the pulsation period, such study works best for long orbital periods and Cepheids with short pulsation periods. Moreover, intrinsic period changes of Cepheids may result in additional erratic shifts which makes the detection of the orbital motion very hard or even impossible.

To improve the detection and the model fit, we tried to fit the anticorrelated binary motion to both components at the same time. To make up for the intrinsic period changes we tried to subtract a varying-order polynomial from the data before the model fit trying to minimize the correlation coefficient, i.e., looking for the highest possible anti-correlation. The results are presented in Fig.~\ref{fig:ltte} and the fitted ephemeris data in Table.~\ref{tab:oc_orb}.

With such a proceduere we eventually detected convincing anti-correlated cyclic phase shifts due to binary motion for five double Cepheids. For SMC-CEP-1526 the binary motion is extremely clear with correlation factor $r=-0.97$ and the orbital period of about 2260 days.  For two other Cepheids (SMC-CEP-2699 and SMC-CEP-3674) the anti-correlated cyclic behavior is less significant but also continues for at least two cycles ($P_\textrm{orb}=$ 2600 and 2400 days). For the first one we had to subtract a polynomial from the data so the amplitudes are affected, while the second exhibits more noisy data. For SMC-CEP-3115 the anti-correlation is clear ($r=-0.94$) but the orbital period is unknown because the orbital cycle has not been covered yet.

The results for the LMC objects using only OGLE data were not conclusive, so we decided to repeat the analysis with the addition of the available MACHO data. For LMC-CEP-0835 this resulted in somewhat distorted O-C data for the secondary but globally the anti-correlated behavior is easy to see. Although a bit more than one cycle is covered and formal errors are small, because of additional intrinsic period changes the accuracy is probably much worse than that (at least two cycles are needed to have more accurate results). Nevertheless the orbital period is probably very long.
Regarding LMC-CEP-0571, the results are similar for OGLE-only and OGLE+MACHO data analysis. In both cases approximately parabolic anti-correlated variability is detected. Unless a very long orbital period is assumed, such O-C results can be better explained with a simple linear period change of opposite signs. Treating this variability as a pure effect of binary motion, yields also unfeasible high Cepheid masses for their pulsation periods. For SMC-CEP-2893 we could not detect any significant LTTE so we expect a shorter orbital period. For the Milky Way objects we do not have enough data and the time span is too short for the O-C analysis.

\begin{figure*}
    \centering
    \includegraphics[width=0.49\linewidth]{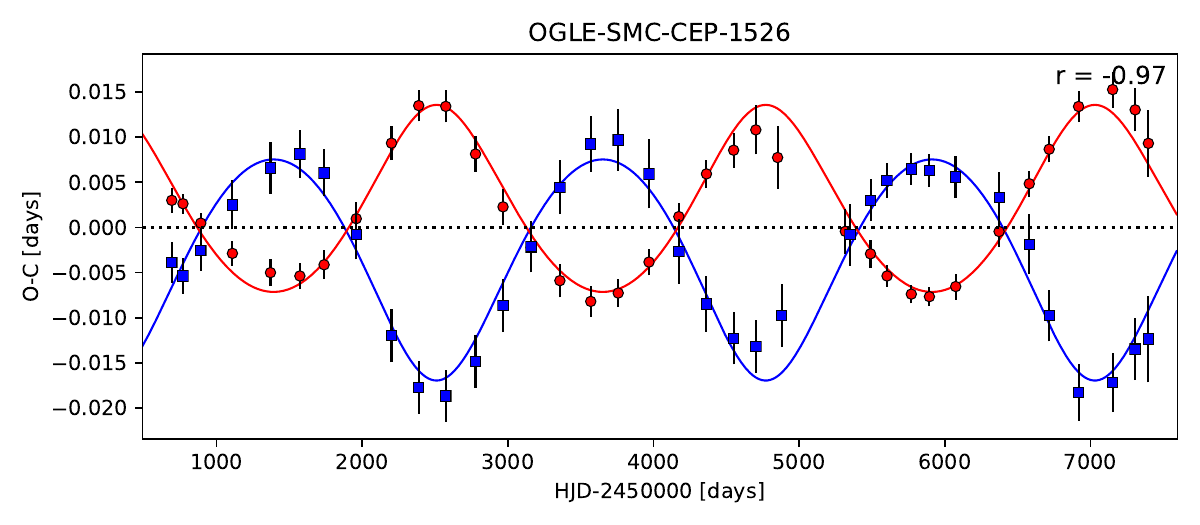}
    \includegraphics[width=0.49\linewidth]{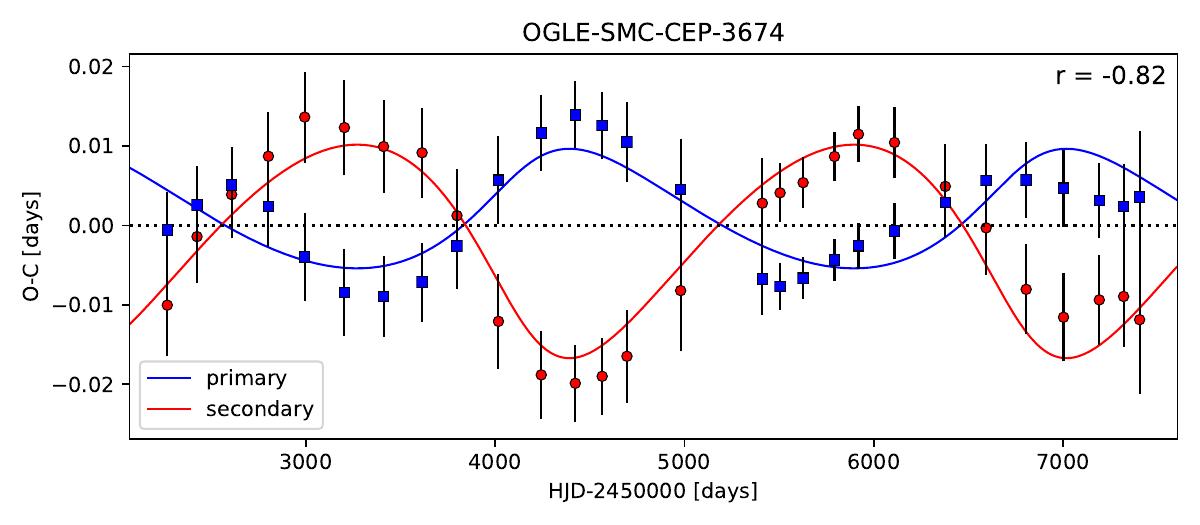}
    \includegraphics[width=0.49\linewidth]{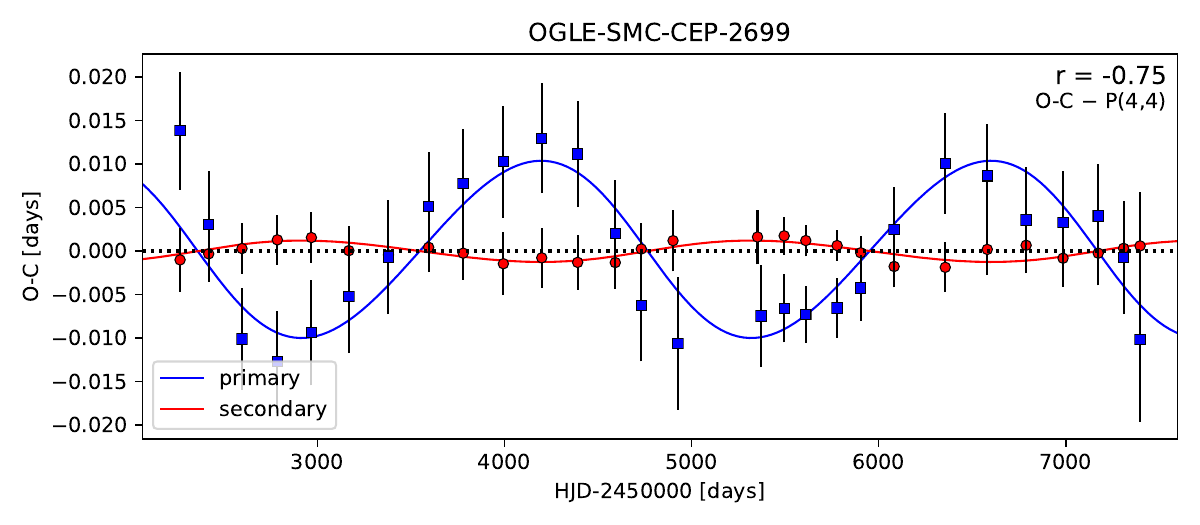}
    \includegraphics[width=0.49\linewidth]{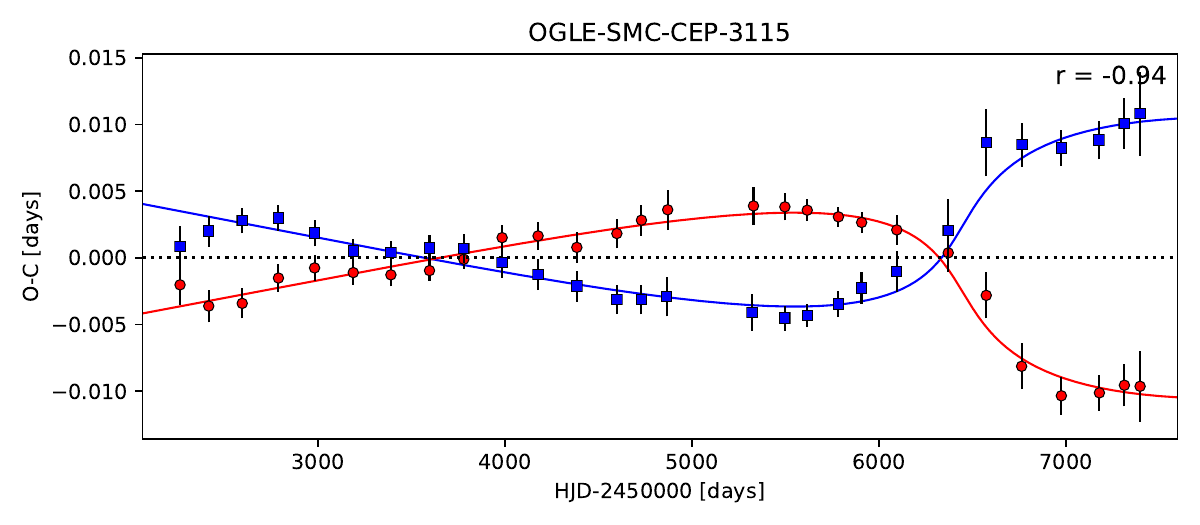}
    \includegraphics[width=0.49\linewidth]{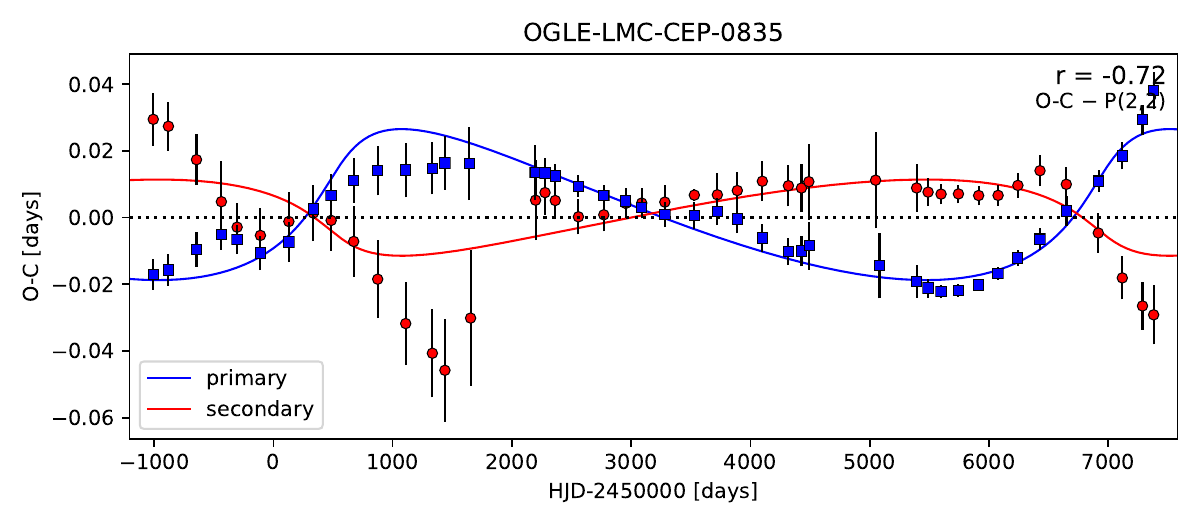}
    \includegraphics[width=0.49\linewidth]{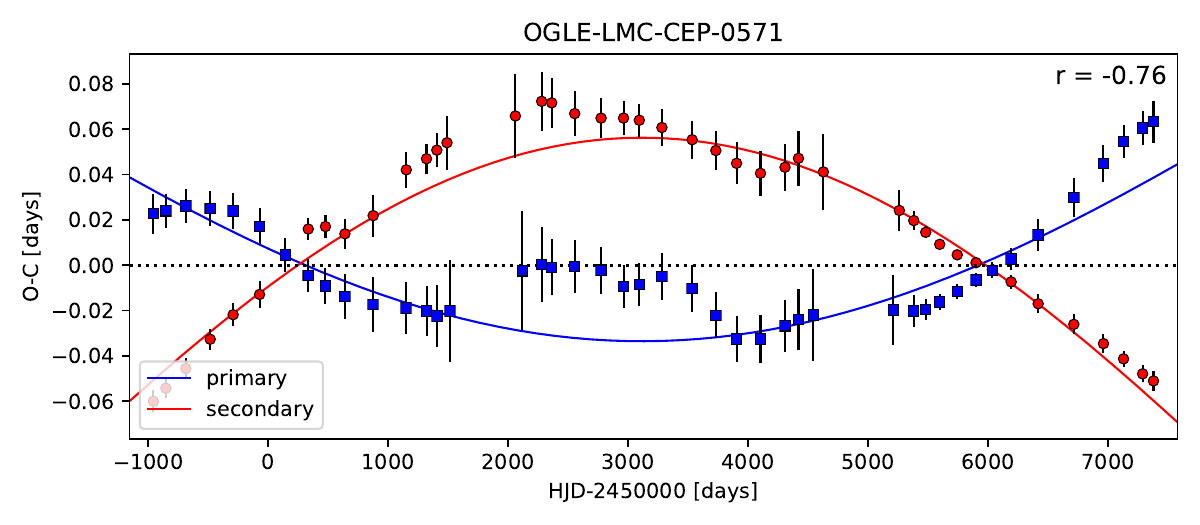}  
    \caption{O-C plots for both components of double Cepheids with anticorrelated phase shifts ($r<-0.7$). Common orbital fit is shown. First four objects show clear LTTE, although for OGLE-LMC-CEP-2699 a 4-order polynomial had to be subtracted from the O-C curves. Anticorrelated variability can also be seen for OGLE-LMC-CEP-0835 (after a subtraction of a parabola) but the secondary shows strong deviation at 1500 days. The phase shifts of OGLE-LMC-CEP-0571 can be explained by anticorrelated linear period change of the components and do not prove binarity.}
    \label{fig:ltte}
\end{figure*}

\begin{table}
\centering
\caption{Orbital ephemeris from the O-C data  \label{tab:oc_orb}}
\begin{tabular}{ccc}
\hline
OGLE ID & $T_{0,\textrm{orb}}$  & $P_\textrm{orb}$ \\
        & [days] & [days]    \\ 
\hline
SMC-CEP-1526 & 5907$\pm$17 & 2260$\pm$12 \\
SMC-CEP-3674 & 4500$\pm$40 & 2630$\pm$70 \\
SMC-CEP-2699 & 4160$\pm$60 & 2400$\pm$80 \\
SMC-CEP-3115 & 8900$\pm$800 & 9300$\pm$2300 \\
LMC-CEP-0835 & 1760$\pm$50 & 6440$\pm$70 \\
\hline
\end{tabular}

\tablefoot{Orbital period ($P_\textrm{orb}$) and reference time ($T_{0,\textrm{orb}}$) for double Cepheids with anitcorrelated variability of the components (marked LTTE in Table~\ref{tab:puls_ephem}). For OGLE-SMC-CEP-3115 the full cycle is not covered and the parameters are highly uncertain.}
\end{table}

\subsection{Spectroscopic analysis}

The spectroscopic analysis of the double Cepheids from the sample was a very challenging task because of three (assuming additional binary motion) independent variabilities involved. At the beginning we did not know which set of lines belong to which Cepheid, and because of line changes (depth, width) along the cycle, it was hard to trace the components from one spectrum to another. Large pulsation amplitudes resulted frequently in an interchange of their positions, i.e., one could not trust that the component with higher velocity is always the same Cepheid. This was further complicated by the lack of information about the orbital phase, period and amplitude, which was adding unknown and variable velocity to all the measurements, including a possible interchange of components also in term of average pulsational velocity. To make the task even harder, the widths of the line profiles and average velocity separations frequently led to a blending of the line profiles. 
Only in the case of three SMC systems with well-determined O-C variability (at least two cycles) we had some advantage from knowing approximate orbital periods and phases.

During a long process of trials and educated guesses we were gradually identifying components and characterizing the underlying orbital motion that corresponds to a given Cepheid in the system. Eventually this led to preliminary orbital solutions presented in Fig.~\ref{fig:rvc}.
The expected orbital periods of most of the systems are longer than 5 years and we had no chance to cover a full orbital cycle for them yet. And to increase accuracy, a bit more than one cycle should be covered, to have at least a marginal overlap in phase.
We will thus continue spectroscopic monitoring of these systems until we will be able to obtain a reliable solution, nevertheless, the analysis of our current data already bring important results.

The most important is the detection of the anticorrelated orbital motion of both components for all double Cepheids from the sample, which ultimately confirms their binarity. For the first eight objects shown in Fig.~\ref{fig:rvc} the orbital velocities change considerably (by 10 km/s or more), a significant curvature of the orbital variability is observed (which helps to constrain models) and/or a velocity reversal in regard to the systemic velocity is observed. 
In cases where orbital periods are known from the O-C analysis, after confirming their consistency with the data they were adopted in the models. These periods will be kept until more precise ones from the spectroscopic analysis are available. For SMC-CEP-2893 for which from the lack of detection in the O-C analysis we expected a short orbital period, indeed the preliminary results suggest a period of about 760 days. This is the only case for which our spectroscopic data span sufficiently wide to cover more than one orbital cycle. For LMC-CEP-0571 a full cycle is close to be covered but the solution is uncertain because of not-optimal phase coverage. Its currently estimated period is the second shortest among the sample which is consistent with the the lack of detection in the O-C analysis (assuming the parabolic variablities originate from period changes). For BLG-CEP-067, GD-CEP-0291 and SMC-CEP-3115 the orbital periods were set to match the available data and do not produce unphysical results (e.g., too high Cepheid masses). 

For the last, ninth object (LMC-CEP-0835) in Fig.~\ref{fig:rvc} from the O-C analysis we expect a very long orbital period ($\sim$6500 days) and small RV variation and indeed, the current RV measurements are consistent with this scenario. 
The orbital motion is clearly seen but because of a short phase range that is covered and small RV changes the model is the least constrained one in the sample. The presented model is one of many that fit the current data, therefore new observations are highly anticipated to constrain the orbital parameters. 

The preliminary orbital parameters and physical properties of BIND Cepheids and their components, including their minimum masses ($M_i\sin^3(i)$, where $i$ is the orbital inclination), are given in Table~\ref{tab:bind_props}. In this table we include also period ratios (for easier comparison with mass ratios) and the radii ratios calculated using the period-mass-radius relation from \citet{allcep_pilecki_2018}. For the first two objects (composed of F-mode Cepheids) period ratios come from Table~\ref{tab:basic}, while for the F+1O-mode Cepheid SMC-CEP-3674 we recalculated it using for fundamentalization equation~2 from \citet{Sziladi_2018_feh_per_prat}, which is more appropriate for deriving radii ratios. However, using for SMC-CEP-3674 the period ratio 0.920 from Table~\ref{tab:basic} we would obtain a similar (within errors) value of $R_2/R_1 = 0.925$.

\begin{table*}
\centering
\caption{Preliminary properties of BIND Cepheids  \label{tab:bind_props}}
\begin{tabular}{c|ccccccc|cc}
\hline
OGLE ID & $T_{0,\textrm{prb}}$   & $P_\textrm{orb}$ & $q$ & $M_1\sin^3(i)$ & $M_2\sin^3(i)$ & $A\sin(i)$  & $e$ & $P_2^F/P_1^F$ & $R_2/R_1$ \\
        & [days]  & [days]    &     & [$M_\odot$]    & [$M_\odot$]    & [$R_\odot$] &     &               & 
     \\
\hline
SMC-CEP-1526 & 8271$\pm$15 & 2260$^*$  & 0.70$\pm$0.09 &  3.6$\pm$0.7  &  2.5$\pm$0.6  & 1320$\pm$90 & 0.42$\pm$0.06 & 0.715 & 0.70$\pm$0.04\\
SMC-CEP-2893 & 9210$\pm$6  & 762$\pm$7 & 0.62$\pm$0.06 & 0.05$\pm$0.01 & 0.03$\pm$0.01 &  154$\pm$9  & 0.57$\pm$0.03 & 0.860 & 0.73$\pm$0.03\\
SMC-CEP-3674 & 9256$\pm$21 & 2625$^*$  & 0.94$\pm$0.13 &  1.7$\pm$0.4  &  1.6$\pm$0.3  & 1190$\pm$80 & 0.31$\pm$0.05 & 0.864$^+$ & 0.89$\pm$0.06\\
\hline
\end{tabular}

\tablefoot{Preliminary properties of BIND Cepheids. Preliminary orbital and physical properties of three BIND Cepheids with best defined orbits. For two of them periods are fixed to the values obtained from the O-C analysis (marked with asterisk). In the penultimate column period ratios are shown for easy comparison with mass ratios. Note that for SMC-CEP-3674 the period ratio (marked with +) was calculated using for fundamentalization equation~2 from \citet{Sziladi_2018_feh_per_prat}. In the last column radii ratios calculated from period-mass-radius relation \citep{allcep_pilecki_2018} are provided.}
\end{table*}

\begin{figure*}
    \centering
    \includegraphics[width=0.42\linewidth]{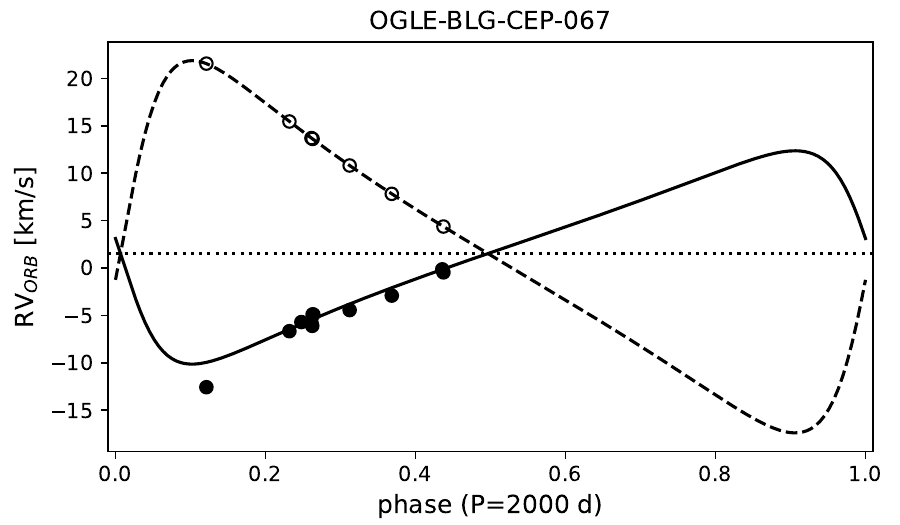}
    \includegraphics[width=0.42\linewidth]{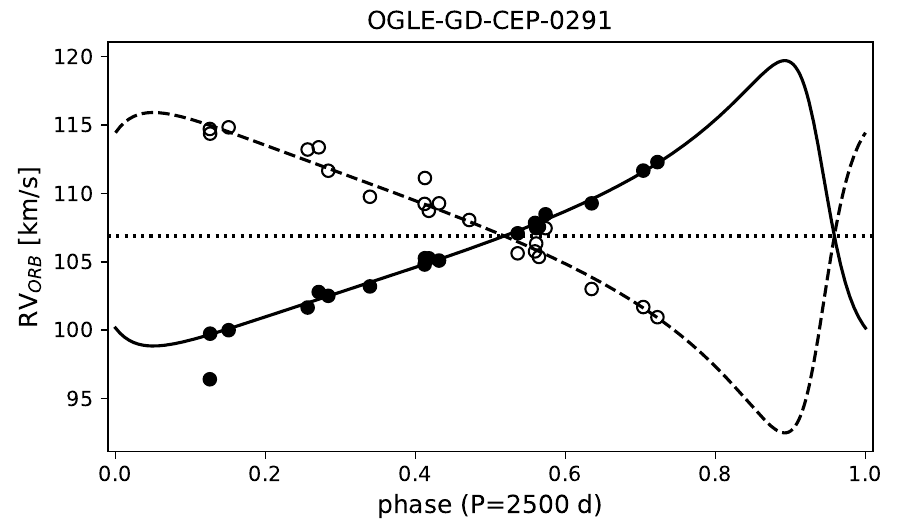}
    \includegraphics[width=0.42\linewidth]{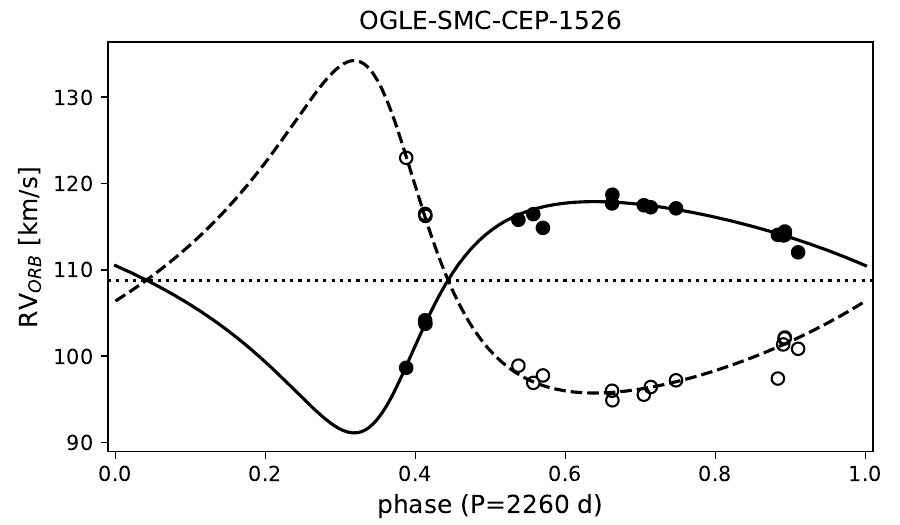}
    \includegraphics[width=0.42\linewidth]{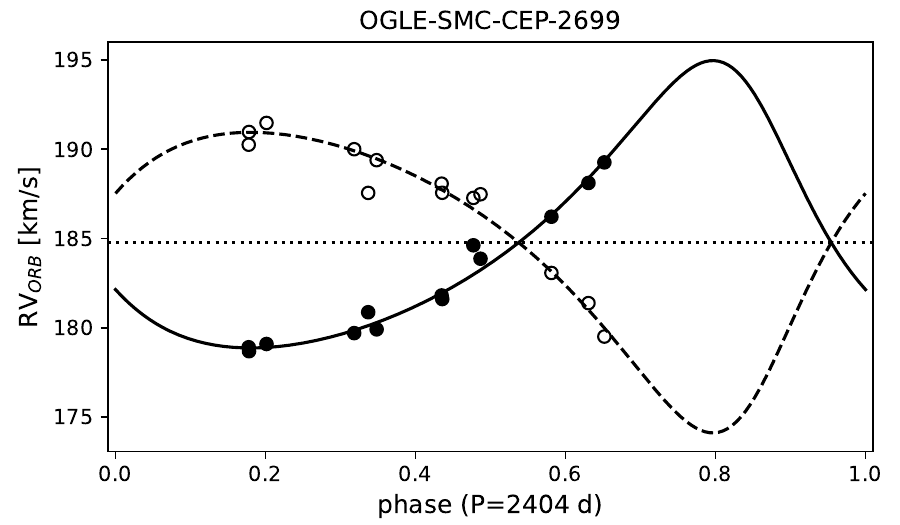}
    \includegraphics[width=0.42\linewidth]{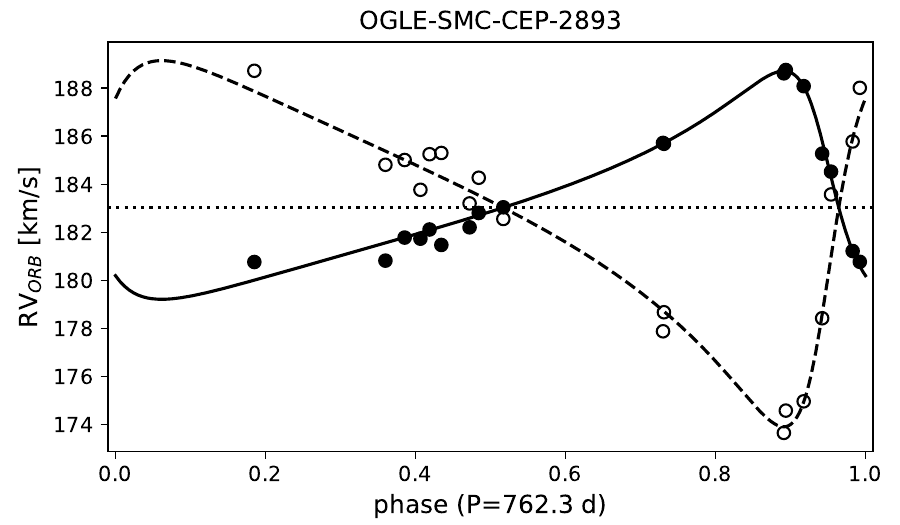}
    \includegraphics[width=0.42\linewidth]{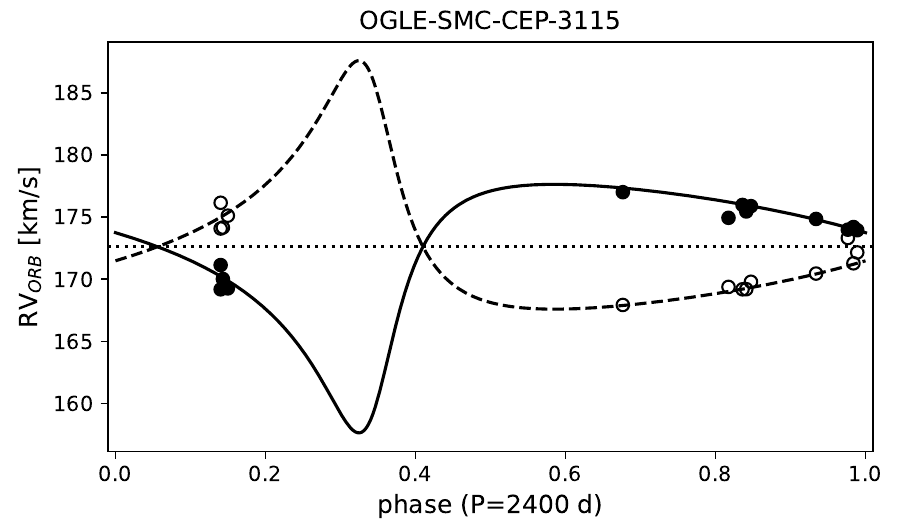}
    \includegraphics[width=0.42\linewidth]{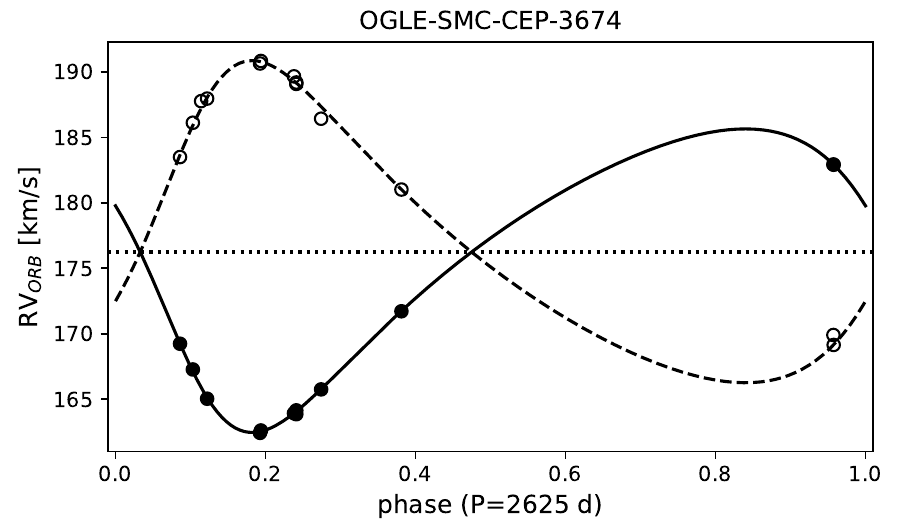}
    \includegraphics[width=0.42\linewidth]{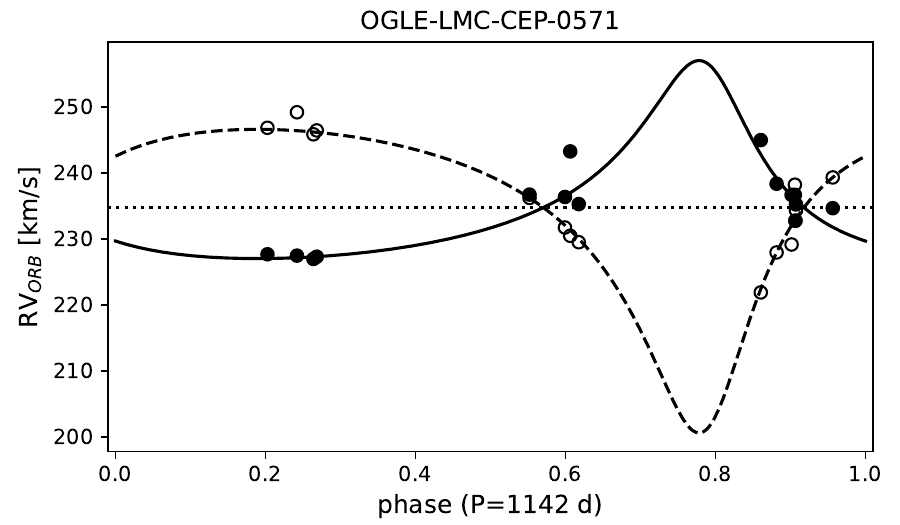}
    \includegraphics[width=0.42\linewidth]{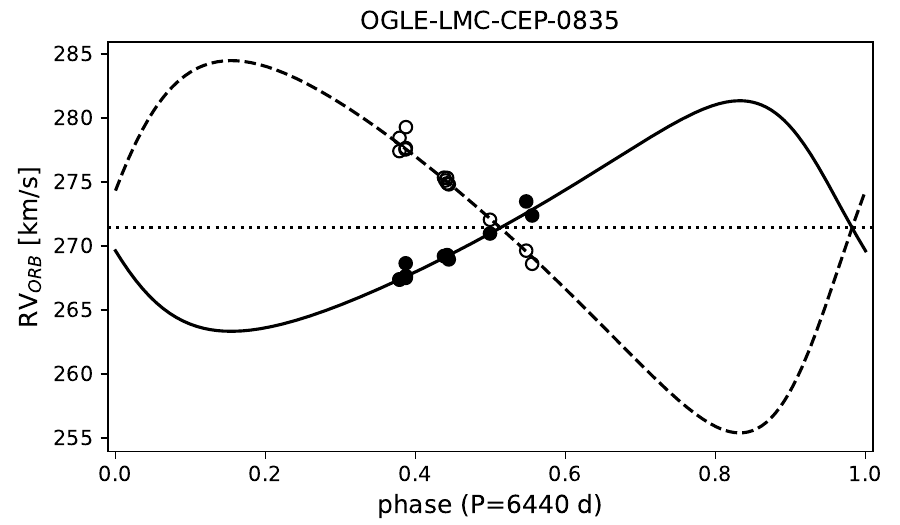}
    \includegraphics[width=0.42\linewidth]{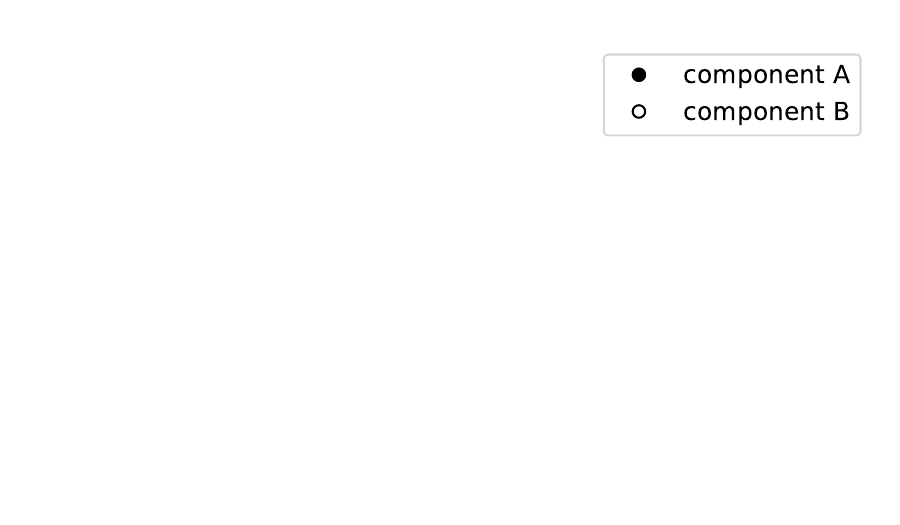}
    \caption{Preliminary orbital radial velocity curves. Only for OGLE-SMC-CEP-2893 and OGLE-LMC-CEP-0571 the orbital periods come from an orbital solution. For three SMC (1526, 2699, 3674) and one LMC (0835) system the period is taken from the LTTE model, while for the rest it is set to roughly match the data and do not produce unphysical results.}
    \label{fig:rvc}
\end{figure*}

For three BIND Cepheids with the best-covered pulsation cycles we show the corresponding RV curves in Fig.~\ref{fig:rvpuls2}. For other systems, the pulsation is taken into account in the model to reduce the scatter but the pulsational RV curves are either approximate (low order Fourier series) or over-fitted (high order Fourier series), depending on what serves better to recover the orbital motion. For example, for objects with steep RV changes the latter works better.

\begin{figure*}
    \centering
    \includegraphics[width=0.49\linewidth]{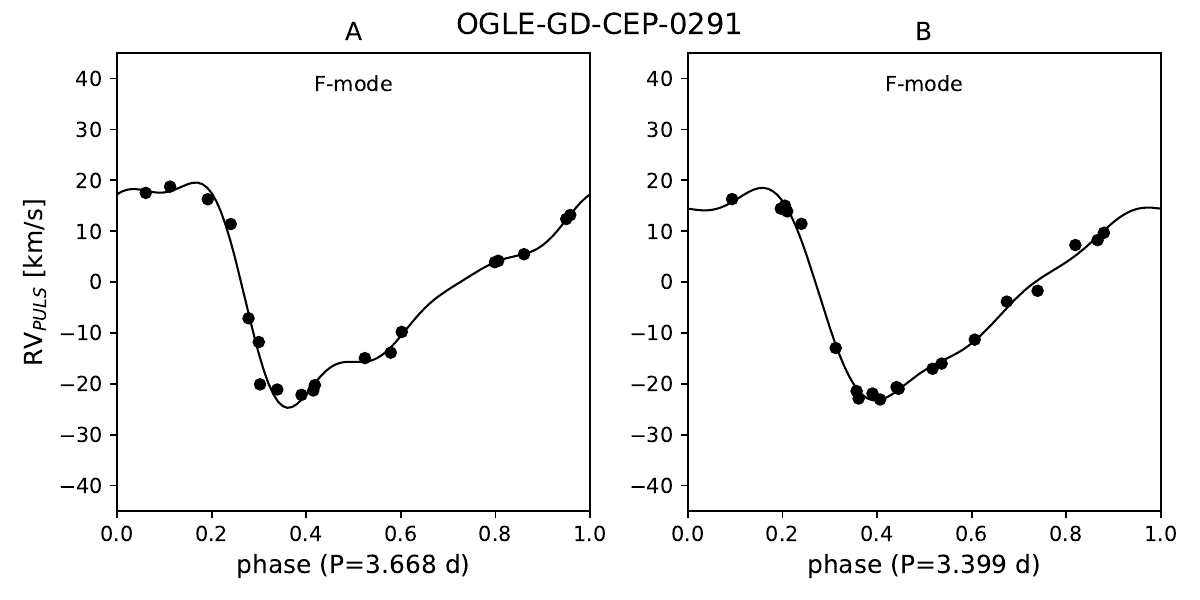}
    \includegraphics[width=0.49\linewidth]{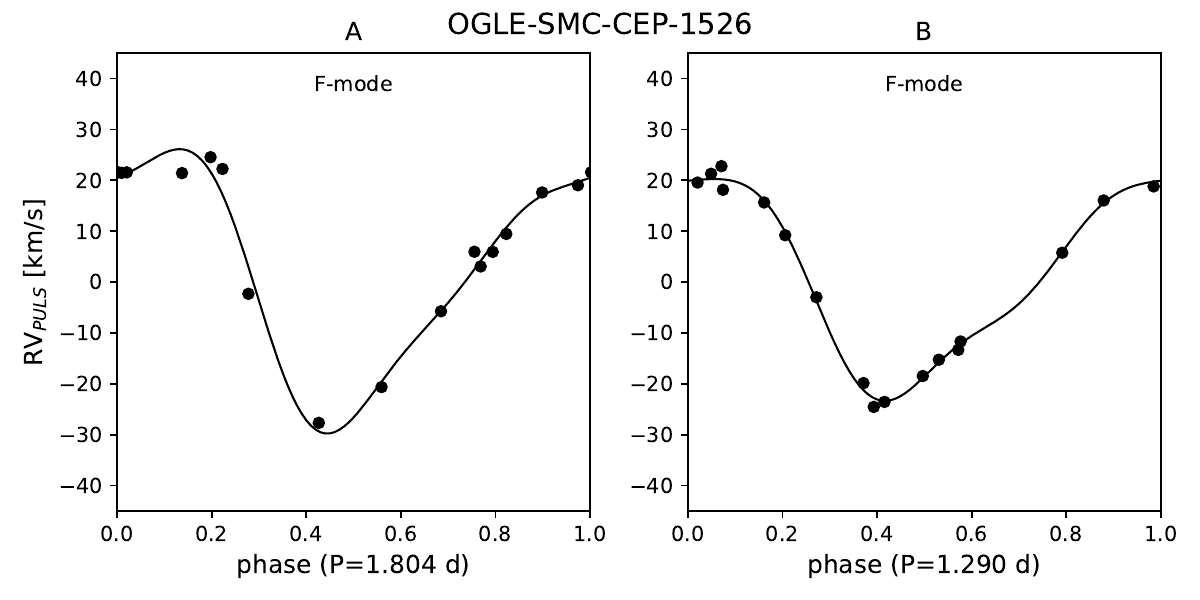}
    \includegraphics[width=0.49\linewidth]{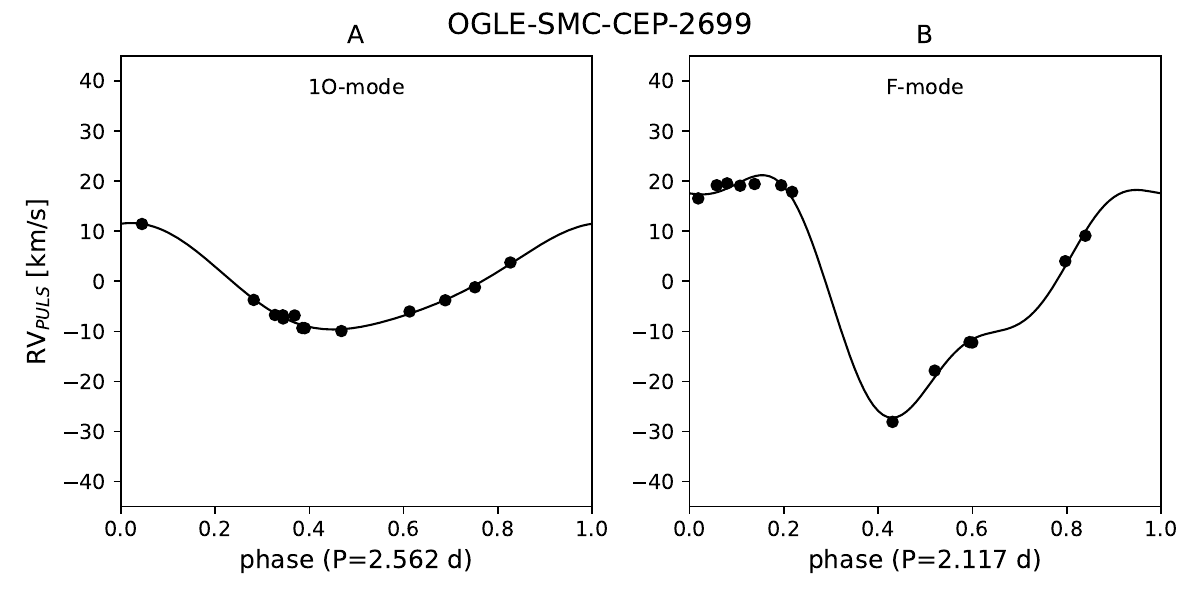}
    \includegraphics[width=0.49\linewidth]{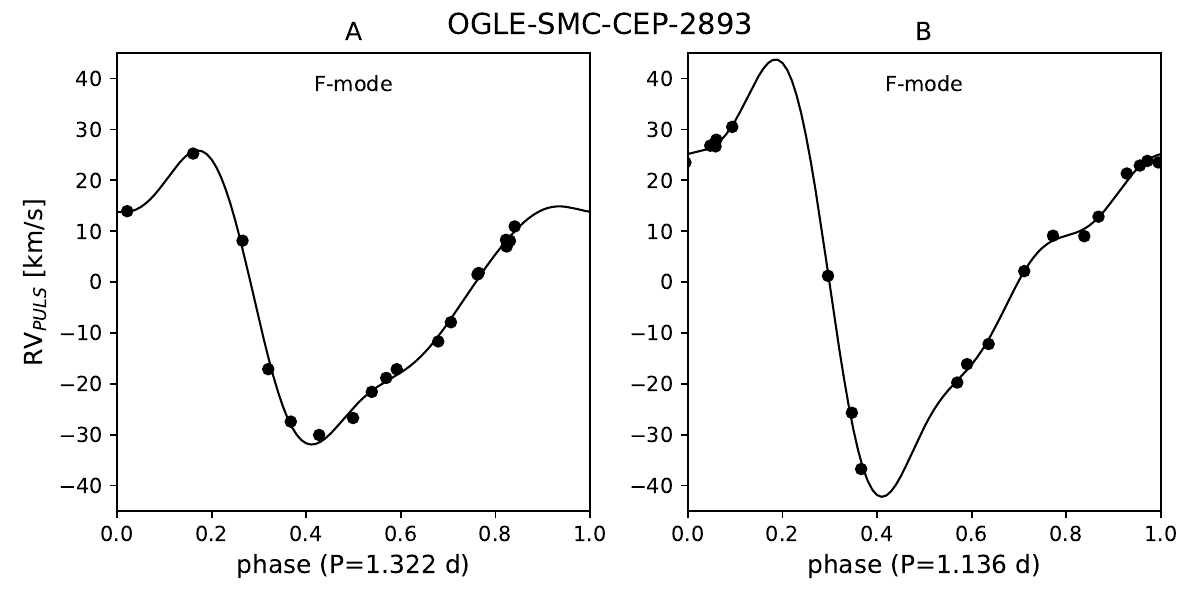}
    \includegraphics[width=0.49\linewidth]{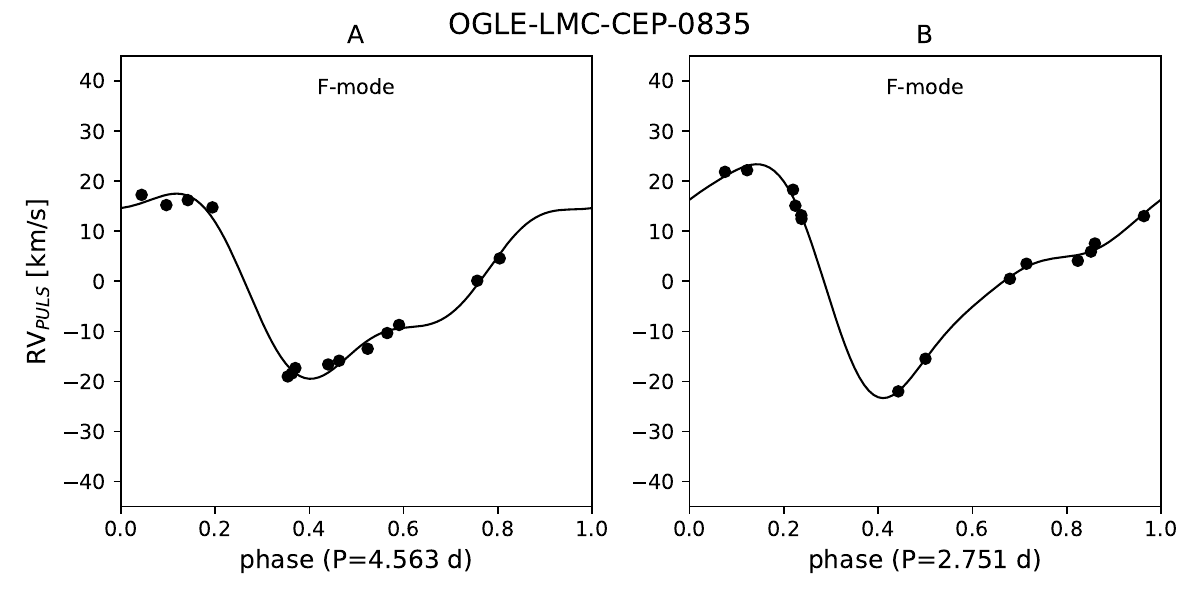}
    \caption{Preliminary pulsational radial velocity curves for five binary double Cepheids with best-covered pulsation cycles. For each system, RVs for both Cepheid components (A and B) are shown. Solid line is a Fourier series fit to the data ($3^{rd}$ to $5^{th}$-order for F mode and $2^{nd}$-order for the 1O Cepheid). The span of the Y-axis is the same for all panels.}
    \label{fig:rvpuls2}
\end{figure*}

\newpage
\subsection{Milky Way sample}

\textbf{ GD-CEP-0291} is composed of two $F$-mode Cepheids. It is the brightest and the closest double Cepheid and also the closest among any type double-lined binary Cepheids known. Because distances from the Gaia mission \citep{GAIA_MISSION_2016} were discrepant (4.3 kpc from GSP-Phot and 11 kpc from parallax; \citealt{Lindegren_2021_astrom_solution}) we calculated a distance to this star using the multiband method \citep{Gieren_2005_multiband_Sculptor}, period-luminosity relations of \citet{Breuval_2022_PLRs} and additional photometric data from the Gaia \citep{GAIA_DR3_2023} and the 2MASS survey \citep{Skrutskie_2006_2MASS}. As our object is a double Cepheid the method had to be slightly modified. From the P-L relations we obtained the expected absolute magnitudes of individual Cepheids and calculated their combined absolute magnitudes. These values were then used as an input to the multiband method together with the apparent magnitudes of the unresolved double Cepheid (see Fig.~\ref{fig:mbdist}, left). Such a procedure yielded a distance of $10.7 \pm 0.6$ kpc, in agreement with the one obtained from the parallax. As a byproduct, the reddening, E(B-V) = $1.00 \pm 0.06$ mag, was also determined. We note that Gaia and OGLE photometry are not consistent. To be conservative, to determine the uncertainties the original photometric errors were thus increased to obtain the reduced $\chi^2$ of about 1.

Because of its wide orbit and closeness (it is located five times closer than the LMC systems), GD-CEP-0291 will be probably the best target for future interferometric observations among all the binary Cepheids. From the current model the expected maximum angular separation is about $1$ $mas$, more than six times higher than for the currently widest orbit of the binary Cepheid LMC-CEP-4506 \citep{cep9009apj2015}.
Once the astrometric orbit is obtained through inteferometry, its combination with the known spectroscopic orbit will directly provide a geometrical distance to the object and the masses of the components.

\textbf{ BLG-CEP-067} is a double Cepheid composed of two $1O$ Cepheids with significantly different periods (see Table~\ref{tab:basic}) that may suggest quite a different components masses. And indeed, the mass ratio obtained from our RV data is consistent with the period ratio, i.e. the shorter-period Cepheid is significantly less massive than the longer-period one. From the two options mentioned at the end of Section~\ref{sec:objects}, this points to the merger scenario and not the combination of a first and a subsequent-crossing Cepheid. Using the multiband method mentioned above we determined the distance to this star $d=26.3 \pm 1.1$ kpc and the reddening E(B-V) = $0.90 \pm 0.04$ mag (Fig.~\ref{fig:mbdist}, right). According to this measurement this system is located far beyond the Galactic bulge, rather in the disk, and only by coincidence is observed close to the Galactic center.

\begin{figure*}
    \centering
    \includegraphics[width=0.4\linewidth]{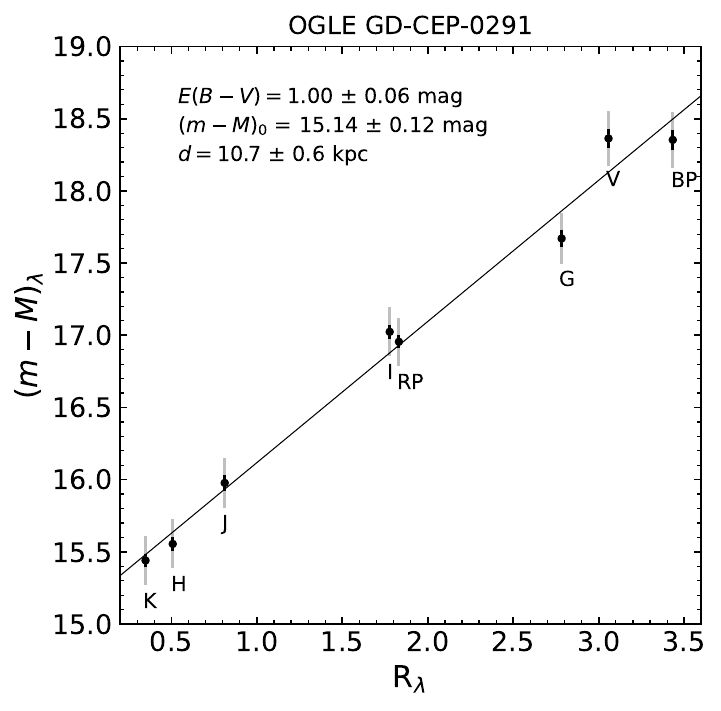}
    \includegraphics[width=0.4\linewidth]{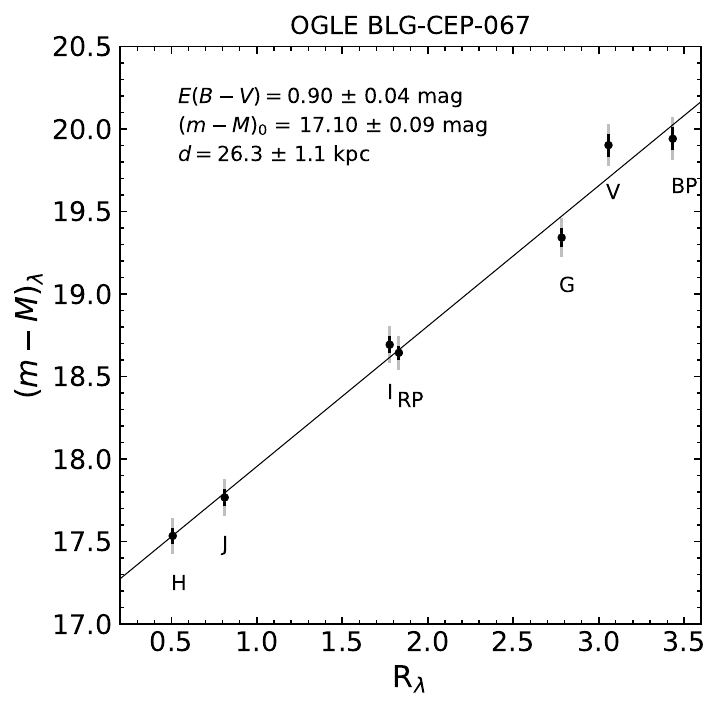}
    \caption{Multiband method applied to the two Milky Way Cepheids providing distances and reddening values. OGLE-GD-CEP-0291 results to be the closest double and, in general, double-lined binary Cepheid known. BLG-CEP-067 is apparently far beyond the Galactic bulge. The original photometric errors are marked in black, while those increased to obtain the reduced $\chi^2=1$ are marked in gray.}
    \label{fig:mbdist}
\end{figure*}

BLG-CEP-067 and GD-CEP-0291 are the first double-lined binaries with a Cepheid (two Cepheids in these cases) in the Milky Way, where both components are giant stars. There are many other MW binary Cepheids known but all of them are composed of a Cepheid and an early-type main-sequence component\footnote{There are some visual pairs of Cepheids, like CE Cas AB, but even if they  eventually result to be gravitationally bound, in practice we would have little gain from that as their orbits would be extremely wide, impeding a meaningful analysis of their orbital motion.}. This opens up a possibility for the first in our galaxy accurate mass determinations from Cepheids in SB2 systems composed of giants.

\subsection{Large Magellanic Cloud}

\textbf{ LMC-CEP-0571} is one of the three mixed-mode ($F$+$1O$) double Cepheids. The period ratio (after fundamentalization of the 1O period) is very close to unity so we expect the same about the mass ratio but we cannot confirm it at the moment as the current solution is very uncertain. Both Cepheids in the system exhibit high period change rates of opposite signs (see Fig.~\ref{fig:ocs}). The preliminary orbital period of this system ($P \sim 1310$ d; we do not have a reliable value from the O-C analysis) is the second shortest in the sample.

\textbf{ LMC-CEP-0835} is composed of two F-mode Cepheids with very low period ratio of about $0.6$. It contains a Cepheid with the longest pulsation period in the sample. Both the O-C analysis and our spectroscopic data confirm it is a binary system with a very long orbital period ($\sim 18$ yr) and small amplitude, which made the observations and analysis very challenging. More data spanning at least a few more years are needed to determine the mass ratio and check why the pulsation periods are so distinct. 

\subsection{Small Magellanic Cloud}

The most numerous group of BIND Cepheids belong to the SMC. They are the first spectroscopically confirmed SB2 systems with at least one Cepheid component in this galaxy.

\textbf{ SMC-CEP-1526} has its orbital ephemeris very well determined from the O-C analysis and thus we can obtain quite reliable preliminary orbital solution with not sufficiently covered orbital RV curve. The ratio of pulsation periods for this F+F double Cepheid ($P_2^F/P_1^F = 0.715$) suggests mass ratio different from unity. And indeed, our spectroscopic data confirms that the secondary (with the shorter period) is significantly less massive than the longer-period primary (see Table~\ref{tab:bind_props}, similarly to BLG-CEP-067 suggesting a merger-origin of the primary component. The preliminary minimum masses (3.6$M_\odot$ and 2.5$M_\odot$) are high and suggest also a high inclination of the orbit.

\textbf{ SMC-CEP-2699} is another mixed-mode ($1O$+$F$) double Cepheid. It has the lowest period ratio in the sample,  $P_2^F/P_1^F = 0.561$. Surprisingly, our preliminary spectroscopic solution indicates the shorter period secondary to be more massive. If this holds true, it would mean that the secondary may not only be of merger origin but also on the first crossing, trying to catch up with the more evolutionary advanced but currently lower-mass primary. The orbital period from the O-C analysis (2400 days) is consistent with the RV curves and is too long for the mass transfer to be the cause of the mass difference.

\textbf{ SMC-CEP-2893} is composed of two F-mode Cepheids, with pulsation periods of the Cepheids among the shortest in the sample. The system has the shortest orbital period ($P\sim 760$ d) among all the BIND Cepheids, which is however longer than that for the previously known eclipsing double Cepheid, LMC-CEP-1718.
This is the only system for which we can determine all the orbital parameters, including orbital period, from the radial velocities alone. The derived minimum masses suggest a very low inclination of the system (it is seen almost face-on, with inclincation below $20^\circ$). The period ratio is moderately lower than unity ($P_2^F/P_1^F = 0.860$), which leaves any option possible but our preliminary mass ratio (0.62) is significantly different from unity, suggesting a merger origin of one of the Cepheids.
However, we are still cautious with this solution because of low RV amplitudes and still poorly covered pulsational RV curves. Moreover, due to the close-to-integer orbital period of 2.09 years, most of the spectra were acquired around two nonoptimal orbital phases ($\sim$0.4 and $\sim$0.9).

\textbf{ SMC-CEP-3115} is also composed of two F-mode Cepheids with very short pulsation periods. According to the O-C data the system has long orbital period and is probably very eccentric. Current RV data span enough to confirm the binary motion but not enough to constrain the eccentricity. The preliminary solution is very uncertain but suggests similar masses of the components, in agreement with the pulsation period ratio (0.926) of the components.

\textbf{ SMC-CEP 3674} is the third mixed-mode ($F$+$1O$) double Cepheid. It shows quite well defined LTTE in the O-C diagram, with the orbital cycle covered about two times. The orbital period from the O-C diagram is consistent with the spectroscopic data. It is the third BIND Cepheid for which a more reliable orbital solution could be obtained. Moderate minimum masses suggest orbital inclination around $45^\circ$. The preliminary mass ratio is $0.94$, which is consistent with the period ratio of $0.920$.

\subsection{Period ratios}\label{sec:perrat}

We mentioned in Section~\ref{sec:objects} that period ratios very different from unity are suspicious and not expected assuming no previous interaction between the components and excluding uncommon combination of a first-crossing Cepheid and a Cepheid on a blue loop. As such we indicated period ratios around 0.6, which is based on our evolutionary and pulsation theory models presented in \citet{Espinoza_2023_IS_LMC}, where such different periods can hardly be obtained for similar mass Cepheids unless one of the Cepheids is still on the first crossing. 
However, to estimate how rare are the period ratios measured for BIND Cepheids a large and empirical comparison sample is necessary. To do that, we looked into the list of 24 Cepheids in the LMC star cluster NGC 1866 \citep{Musella_2016_ceph_ngc1866}, which should all roughly have similar ages. We calculated all possible combinations of period ratios for these Cepheids, excluding two with uncertain membership and fundamentalizing periods for first-overtone Cepheids.
A histogram of these period ratios compared with those for BIND Cepheids is shown in Fig.~\ref{fig:perrat} (left). As can be seen there, $P_2^F/P_1^F$ values for the NGC 1866 Cepheids lower than 0.8 are very rare and below 0.7 absolutely non-existent. 

However, Cepheids in NGC 1866 have fundamentalized periods between 2.64 and 3.52 days, while $P^F$ of our BIND Cepheids ranges from 1.14 to 4.56 days. To increase the size of the comparison sample and the range of periods we added to the analysis 12 Cepheids ($P^F$ from 2.66 to 4.43 days)  from another LMC cluster, NGC 2031 \citep{Bertelli_1993_NGC1866_2031_Cepheids}. A histogram for the combined datasets is shown in the right panel of the same figure. According to this test, taking randomly a pair of Cepheids from any of the two clusters one would have little chance ($\sim$7\%) of obtaining a ratio lower than 0.8, and only about 2\% for ratios below 0.7. On the contrary, half of BIND Cepheids have these values lower than 0.8, and 30\% of them values below 0.7.

Taking into account that the Cepheid membership for NGC~2031 is not as certain as for NGC~1866 we made a test removing different NGC~2031 Cepheids from the list and repeating the analysis. For such variants, chances varied from 4 to 8\% for period ratios lower than 0.8 and from 0.3 to 2\% for ratios lower than 0.7, meaning it would be hard to increase these values considerably above those determined in the previous paragraph. We also note that for the visual pair of Cepheids, CAab Cas (which is likely gravitationally bound; see \citealt{Kervella_2019_Multiplicity}), the period ratio is 0.87 \citep{Opal_1988_CE_Cas_double_Cepheid}.

We can compare these results with what we know about eclipsing binary double Cepheid, LMC-CEP-1718. For that Cepheid the period ratio is 0.786. Although the mass ratio is close to unity, the slightly less massive Cepheid is more luminous and seems to be more evolutionary advanced, which points to some kind of disturbance in the past evolution of this double Cepheid. This can be a border case, where the components passed through a weak interaction, which did not change the mass ratio strongly. Note that this is the tightest system among all the BIND Cepheids with the orbital period of 413 days.

\begin{figure*}
    \centering
    \includegraphics[width=0.49\linewidth]{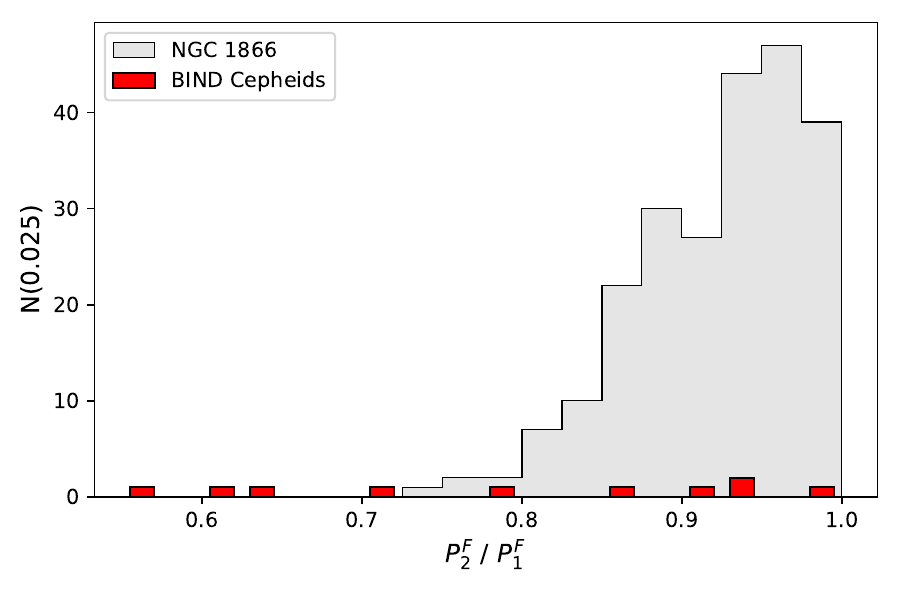}
    \includegraphics[width=0.49\linewidth]{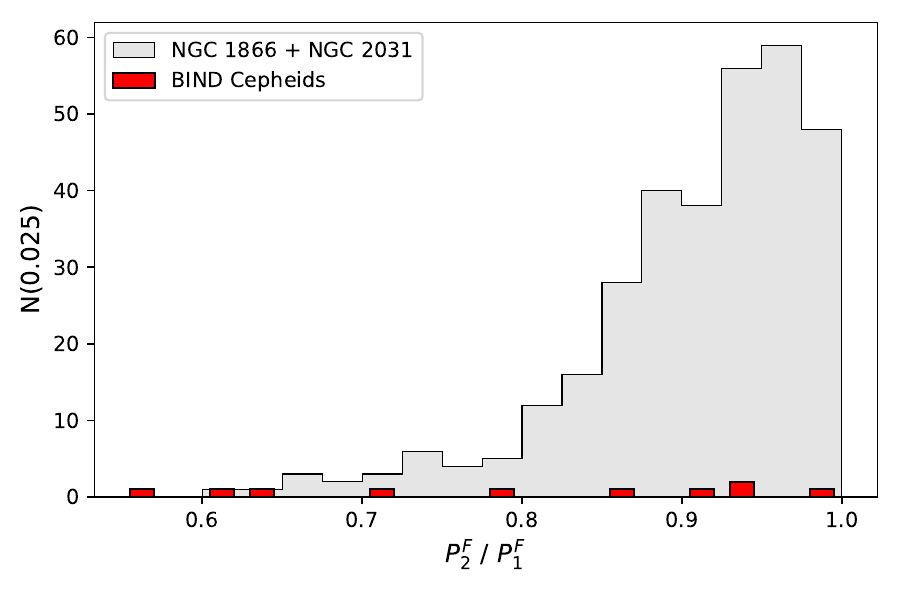}
    \caption{Histogram of period ratios for all combinations of Cepheids in NGC 1866 and (separately) in NGC 2031 compared with period ratios of BIND Cepheids. Only for $\sim$7\% of these combinations period ratios are lower than 0.8, while period ratios of 50\% of BIND Cepheids are below this limit.}
    \label{fig:perrat}
\end{figure*}

\newpage
\subsection{Period-Luminosity relations}\label{sec:pl}

Most of the Cepheids are expected to exist in binary systems, with the great majority of them having early-type main sequence companions that are much fainter and are hardly seen in the spectra \citep{BohmVitense_1985_ApJ_cep_blue_companions,Kervella_2019_Multiplicity}. As the systems we analyzed here are known to be composed of two Cepheids there is a tempting possibility to check the effect of binarity on the average brightening of the period-luminosity (P-L) relations. 

The presence of two Cepheids allows us to determine the luminosity ratio from the known period ratios (see Table~\ref{tab:basic}). This in turn allows us to split the total observed flux between the two components. We did this for their V and I-band magnitudes and calculated corresponding Wesenheit magnitudes, $W(V,I)_i=I-1.55(V-I)$ for individual Cepheids. With the same approach as in P21, we prepared the P-L relations for all Cepheids in their host galaxies, fundamentalizing periods of $1O$ Cepheids.
We then compared the $W(V,I)_i$ values with these relations and found that practically all the individual components lie significantly below them, sometimes on it, and never significantly above. For the LMC systems the average difference is $0.016 \pm 0.012$ mag, while for the SMC it is $0.053 \pm 0.022$ mag. A weighted mean of both is $0.024 \pm 0.010$ mag. Although this detection is not very firm, the sign is what we expect and the value looks reasonable. This is also the first empirical determination of P-L relation shift due to binarity.

One may note, that this is only a shift due to the presence of secondary components, while there may exist higher-order multiple systems with Cepheids. However, we can expect the influence of higher-order components to be much lower because: 1) they are more rare, 2) they are on average less massive and luminous, and 3) they would have less significant relative effect on the total brightness.

\section{Conclusions and perspectives}\label{sec:conclusions}

\citet{Alcock_1995_double_cepheids} identified first three double Cepheids in the LMC galaxy (identified LMC-CEP-0571,0835,1718 in this work). Although they speculated they could be binaries, the conclusion was that the physical connection between them could not be determined without a long time scale spectroscopic monitoring. They expected periods longer than 1.5 years.
In 2014 we performed the spectroscopic study of one of these systems (LMC-CEP-1718; \citealt{cep1718apj2014}) revealing an orbital period of 1.13 yr. This was the first, and for a long time the only, double Cepheid spectroscopically confirmed as a binary system (i.e., a BIND Cepheid).

In our current study we confirmed nine more double Cepheids from the MW, LMC and the SMC galaxies to be components of binary systems, increasing the total number of BIND Cepheids from 1 to 10. With this discovery we also tripled the number of all spectroscopically confirmed Cepheids in double-lined binary systems.
The SMC and Milky Way BIND Cepheids are the first found in double-lined systems in their host galaxies. The MW ones also also the closest to us of such type. Located at a distance of around 11 and 26 kpc, they are 5 and 2 times closer than previously known SB2 Cepheids in the LMC.
As there are no more other known unresolved double Cepheids than those presented here, these ten BIND Cepheids will be probably all we know for quite a long time.

As expected from the O-C analysis, the orbital periods of seven systems (or eight counting LMC-CEP-0571 with a lower limit of 3.6 yr) are indeed long, longer than 5 years. This makes the probability of any of these systems to exhibit eclipses very low. In this sense we are very lucky that the shortest period BIND Cepheid (LMC-CEP-1718; \citealt{allcep_pilecki_2018}) resulted to be eclipsing even if the eclipse is grazing, as there was only about 8\% chance for that, and the chance for eclipses in other BIND Cepheids is much lower still -- about 3\% on average. In our current sample there is only one system with relatively short period, SMC-CEP-2893 but as mentioned before, it is probably positioned almost face on. 
Looking from the other side, the longest orbital period of about 18 years was found for LMC-CEP-0835 basing on the O-C analysis while confirming the binarity spectroscopically. This is the longest period binary confirmed in another galaxy but such long periods, or even longer, are quite common in the Milky Way, also among binary Cepheids (see e.g. a recent work of \citealt{Cseh_2023_binCep_V1344Aql}).

Our study shows that the O-C analysis provides quite reliable results regarding detection of binarity and orbital periods whenever more than one cycle of the orbital motion is covered. In case of BIND Cepheids, the anticorrelated O-C behavior may also indicate binarity even when the cycle is not covered completely.

For the first time we empirically estimated an effect of binarity on period-luminosity relations. The sign of the shift is as expected and the value, $\Delta W(V,I)=0.024 \pm 0.010$ mag, looks reasonable. A further, more detailed study will be however needed to improve and confirm this value.

\subsection{Perspectives}

This is a long-term project and with this paper we are finishing its first phase whose goal was to spectroscopically confirm binarity of all known double Cepheids and obtain first estimates of the basic parameters such as orbital periods. With this part behind we may look forward and prepare for the next phase of the project which is to obtain final physical properties of the components and precise orbital parameters of the systems. For that we are planning to continue monitoring of all the objects until the number of data is sufficient for a reliable solution, publishing final results for a given system or systems as soon as they are ready.

Unless we detect eclipses for any of these systems, a direct measurement of masses will not be possible. From the orbital solution measuring only mass ratios and $M\sin(i)^3$ is possible because of unknown inclination ($i$), so we will have only lower limits. However, from the directly measured mass ratios and the extremely precise period ratios readily available from the photometric data, radii ratios can also be calculated using e.g. period-mass-radius relation \citep{Bono_2001APJ_cep_MR_mass_discrep,allcep_pilecki_2018} or basic pulsation theory models \citep{t2cep098apj2017,sm08}.
Using additional data, like a known distance (individual or to the host galaxy in the case of the LMC and SMC objects), observed brightness and temperature measured from spectra, absolute radii and masses can also be determined. This means that potentially we may have new accurate mass estimates from SB2 systems for up to 18 Cepheids, which will be a huge improvement comparing with our current knowledge (6 direct measurements and 5 uncertain estimates from SB1 systems; \citealt{allcep_pilecki_2018, Evans_2018_V350SGr_mass}). We estimate the precision of masses obtained this way to be better than 10\%. Together with already published LMC-CEP-1718, this set of nine new BIND Cepheids (10 systems, 20 Cepheids in total) will form an important base for various follow-up studies (e.g., statistical analysis, evolutionary and pulsation modeling).

A very tempting possibility that opens up with the presented BIND Cepheids is to try to resolve them interferometrically and measure the angular separation ($\phi$). Once this is done, it is possible to directly calculate a geometrical distance using a simple formula, d[pc] = 9.2984 $\times$ L[R$_\odot$]/$\phi$[mas], where L is the linear projected separation from the spectroscopic solution. 
As mentioned before, we can start with the closest system GD-CEP-0291, but eventually move to the Magellanic Clouds, using the advantages of BIND Cepheids of being luminous and having large orbital separation of the components which is necessary to resolve them.
Normally, systems with such properties are hard to identify in other galaxies. Some were detected through eclipses during even decades-long microlensing surveys (OGLE, MACHO) but apart from not being efficient, this method favors tighter orbits. The longest orbital period measured so far for an extragalactic system is P=1550 days ($\sim$4 yr; \citealt{cep9009apj2015}), which translates to 0.16 $mas$ separation. 
Therefore, BIND Cepheids are currently our best candidates for a direct and accurate geometric distance determination to the LMC and SMC galaxies, which would eventually lead to the ultimate calibration of the first rung of the cosmic distance ladder.

\begin{acknowledgements}
We thank the referee, L{\'a}szl{\'o} Szabados, for constructive comments on the manuscript.
The research leading to these results has received funding from the Polish National Science Center grant SONATA BIS 2020/38/E/ST9/00486. We also acknowledge financial support from the UniverScale grant financed by the European Unions Horizon 2020 research and innovation programme under the grant agreement number 951549. A.G. acknowledges the support of the Agencia Nacional de Investigaci\'on Cient\'ifica y Desarrollo (ANID) through the FONDECYT Regular grant 1241073.

This paper utilizes public domain data obtained by the MACHO Project, jointly funded by the US Department of Energy through the University of California, Lawrence Livermore National Laboratory under contract No. W-7405-Eng-48, by the National Science Foundation through the Center for Particle Astrophysics of the University of California under cooperative agreement AST-8809616, and by the Mount Stromlo and Siding Spring Observatory, part of the Australian National University.

This work is based on observations collected at the Las Campanas Observatory and the European Southern Observatory (ESO programmes: 106.21GB, 108.229Z, 108.22BS, 110.2434, 110.2436). We thank Carnegie, and the CNTAC (program IDs: CN2021A-36, CN2022A-26, CN2023A-53) for the allocation of observing time for this project. 

This research has made use of NASA's Astrophysics Data System Service.
\end{acknowledgements}

\bibliography{ccbincand2}{}
\bibliographystyle{aa}

\end{document}